\documentclass[preprint,showpacs,preprintnumbers,amsmath,amssymb,showkeys]{revtex4}
\usepackage{graphicx}% Include figure files
\usepackage{dcolumn}% Align table columns on decimal point
\usepackage{bm}% bold math

\begin{document}

\preprint{}

\title{Construction of a polarization insensitive lens from a quasi-isotropic metamaterial slab}
\author{Hailu Luo}
\email{hailuluo@gmail.com}
%\altaffiliation[ ]{}%Lines break automatically or can be forced with \\
\author{Zhongzhou Ren}
\author{Weixing Shu}
\author{Fei Li}
%%\email{Second.Author@institution.edu}
\affiliation{Department of Physics, Nanjing University, Nanjing
210008, China}
\date{\today}% It is always \today, today,

\begin{abstract}
We propose to employ the quasiisotropic metamaterial (QIMM) slab
to construct a polarization insensitive lens, in which both E- and
H-polarized waves exhibit the same refocusing effect. For shallow
incident angles, the QIMM slab will provide some degree of
refocusing in the same manner as an isotropic negative index
material. The refocusing effect allows us to introduce the ideas
of paraxial beam focusing and phase compensation by the QIMM slab.
On the basis of angular spectrum representation, a formalism
describing paraxial beams propagating through a QIMM slab is
presented. Because of the negative phase velocity in the QIMM
slab, the inverse Gouy phase shift and the negative Rayleigh
length of paraxial Gaussian beam are proposed. We find that the
phase difference caused by the Gouy phase shift in vacuum can be
compensated by that caused by the inverse Gouy phase shift in the
QIMM slab. If certain matching conditions are satisfied, the
intensity and phase distributions at object plane can be
completely reconstructed at image plane. Our simulation results
show that the superlensing effect with subwavelength image
resolution could be achieved in the form of a QIMM slab.

\end{abstract}

%Use showkeys class option if keyword
\maketitle

\section{Introduction}\label{Introduction}
About forty years ago, Veselago firstly introduced the concept of
left-handed material (LHM) in which both the permittivity
$\varepsilon$ and the permeability $\mu$  are
negative~\cite{Veselago1968}. He predicted that LHM would have
unique and potentially interesting properties, such as the
negative refraction index, the reversed Doppler shift and the
backward Cerenkov radiation. Veselago pointed out that
electromagnetic waves incident on a planar interface between a
regular material and a LHM will undergo negative refraction. Hence
a LHM planar slab can act as a lens and focus waves from a point
source.  LHM did not receive much attention as it only existed in
a conceptual form. After the first experimental observation of
negative refraction using a metamaterial composed of split ring
resonators (SRRs)~\cite{Smith2000,Shelby2001}, the study of such
materials has received increasing attention over the last few
years. While negative refraction is most easily visualized in an
isotropic
metamaterial~\cite{Smith2000,Shelby2001,Pacheco2002,Parazzoli2003,Houck2003},
negative refraction can also be realized in photonic
crystals~\cite{Notomi2000,Luo2002a,Luo2002b,Li2003} and
anisotropic
metamaterials~\cite{Lindell2001,Hu2002,Smith2003,Zhang2003,Zhou2003,Luo2005,Thomas2005,Luo2006a,Grzegorczyk2006}
have also been reported.

Recently, Pendry  extended Veslago's analysis and further
predicted that a LHM slab can amplify evanescent waves and thus
behaves like a perfect lens~\cite{Pendry2000}. He proposed that
the amplitudes of evanescent waves from a near-field object could
be restored at its image. Therefore, the spatial resolution of the
superlens can overcome the diffraction limit of conventional
imaging systems and reach the subwavelength scale. The great
research interests were initiated by the revolutionary concept.
More recently, the anisotropic metamaterials have been proved to
be good candidates for slab lens
application~\cite{Smith2004a,Smith2004b,Parazzoli2004,Dumelow2005}.
Although the focusing is imperfect, the substantial field
intensity enhancement can readily be observed. In these cases, the
anisotropic metamaterials under consideration are characterized by
a hyperboloid dispersion relation, and the focusing is restricted
to either E- or H-polarized radiation. The recent development in
quasiisotropic metamaterial (QIMM) offers us further opportunities
to extend the previous work and further predict that both E- and
H-polarized waves can be refocused.

The main purpose of the present work is to construct a
polarization insensitive lens by a QIMM slab. For shallow incident
angles the QIMM slab will provide some degree of refocusing in the
same manner as an isotropic LHM slab. We are particularly
interested in exploiting  polarization insensitive effect. First,
starting from the representation of plane-wave angular spectrum,
we derive the propagation of paraxial beams in the QIMM slab. Our
formalism permits us to introduce ideas for beam focusing and
phase compensation of paraxial beams by using the QIMM slab. Next
we want to introduce the inverse Gouy phase shift and negative
Rayleigh length when waves propagating in the QIMM slab. As an
example, we obtain the analytical description for a Gaussian beam
propagating through a QIMM slab. We find that the phase difference
caused by the Gouy phase shift in vacuum can be compensated by
that caused by the inverse Gouy phase shift in the QIMM slab. If
certain matching conditions are satisfied, the intensity and phase
distributions at object plane can be completely reconstructed at
the image plane. Finally, we will discuss what happen when the
evanescent wave transmission through the QIMM slab.

\section{Polarization insensitive metamaterial}\label{SecI}
Before we consider the polarization insensitive lens, we first
analyze what is the QIMM. For anisotropic materials, one or both
of the permittivity and permeability are second-rank tensors. In
the following we assume that both the permittivity and
permeability tensors are simultaneously diagonalizable:
\begin{eqnarray}
\boldsymbol{\varepsilon}=\left[
\begin{array}{ccc}
\varepsilon_x (\omega)  &0 &0 \\
0 & \varepsilon_y (\omega)  &0\\
0 &0 & \varepsilon_z (\omega)
\end{array}
\right],~~~\boldsymbol{\mu}=\left[
\begin{array}{ccc}
\mu_x (\omega)  &0 &0 \\
0 & \mu_y(\omega) &0\\
0 &0 & \mu_z (\omega)
\end{array}
\right],\label{matrix}
\end{eqnarray}
where $\varepsilon_j$ and $\mu_j$  are the relative permittivity
and permeability constants in the principal coordinate system
($j=x,y,z$). It should be noted that the real anisotropic
metamaterial constructed by SRRs is highly dispersive, both in
spatial sense and frequency
sense~\cite{Smith2003,Smith2004a,Smith2004b,Thomas2005}. So these
relative values are functions of the angle frequency $\omega$.

Following the standard procedure, we consider a monochromatic
electromagnetic field ${\bf E}({\bf r},t) = Re [{\bf E}({\bf
r})\exp(-i\omega t)]$ and ${\bf B}({\bf r},t) = Re [{\bf B}({\bf
r})\exp(-i\omega t)]$ of angular frequency $\omega$ incident from
vacuum into the anisotropic metamaterial. The field can be
described by Maxwell's equations~\cite{Chen1983}
\begin{eqnarray}
\nabla\times {\bf E} &=& - \frac{\partial {\bf B}}{\partial t},
~~~{\bf B} = \mu_0 \boldsymbol{\mu}\cdot{\bf H},\nonumber\\
\nabla\times {\bf H} &=&  \frac{\partial {\bf D}}{\partial
t},~~~~~{\bf D} =\varepsilon_0 \boldsymbol{\varepsilon} \cdot {\bf
E}. \label{maxwell}
\end{eqnarray}
The previous Maxwell's equations can be combined in a
straightforward way to obtain the well-known equation for the
complex amplitude of the electric field, which reads
\begin{equation}
\nabla\times( \boldsymbol{\mu}^{-1}\cdot\nabla\times{\bf
E})+\frac{1}{c^2}\frac{\partial^2 {\bf D}}{\partial t^2}
=0,\label{mpe}
\end{equation}
where $c$ is the speed of light in vacuum.

\begin{figure}
\includegraphics[width=6cm]{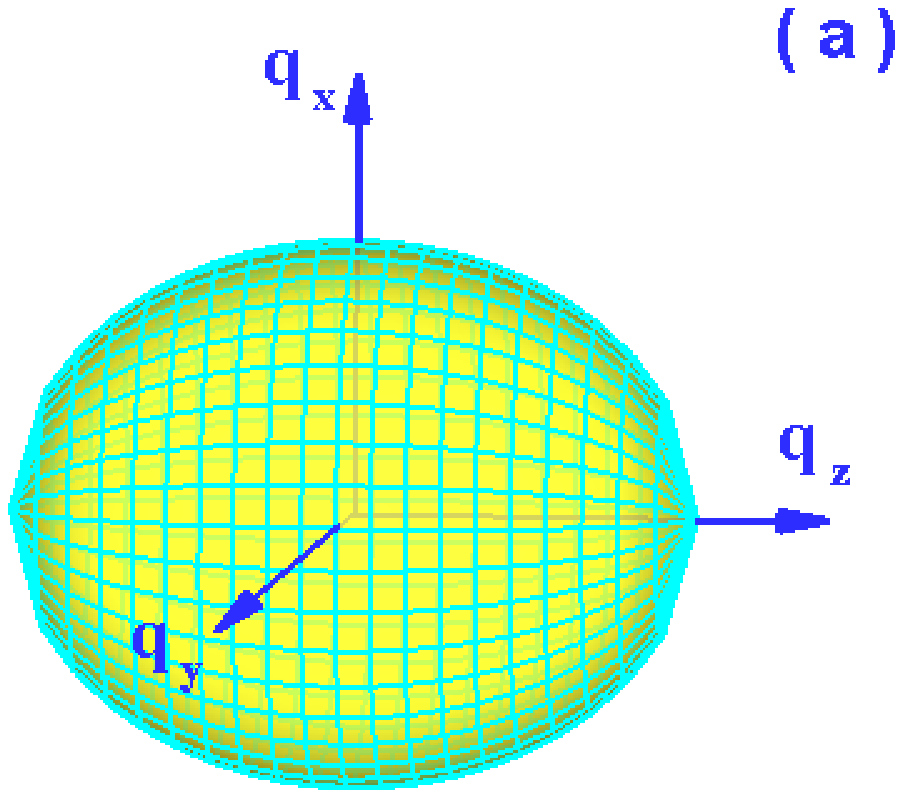}~~~~~
\includegraphics[width=6cm]{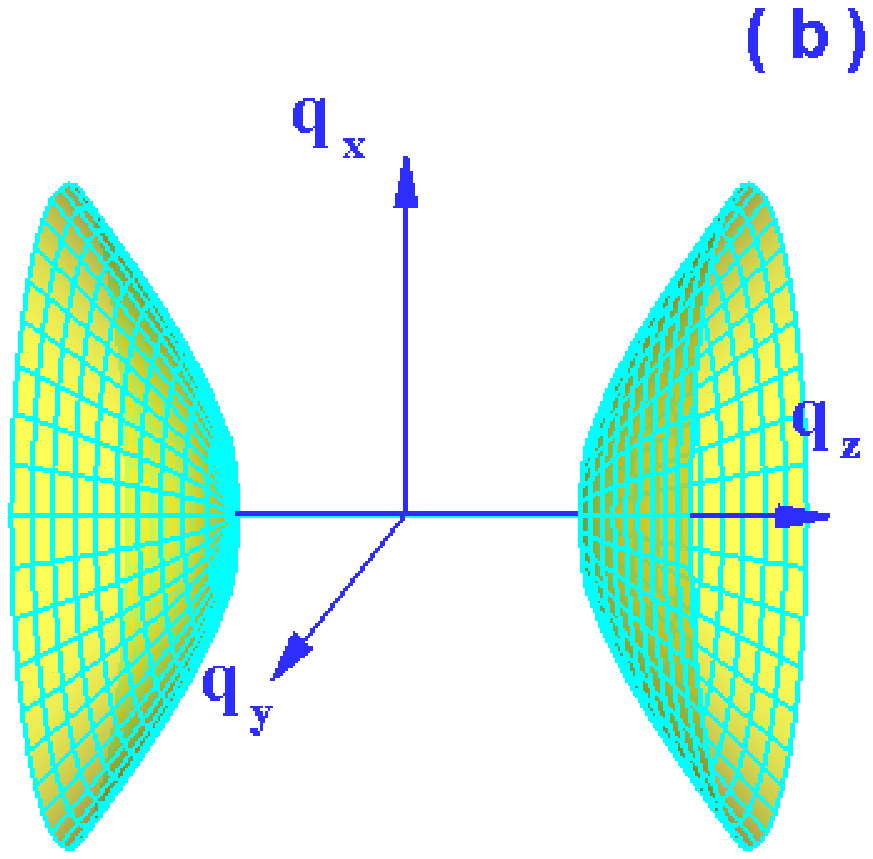}
% Here is how to import EPS art
\caption{\label{Fig1} (Color online) (a) The QIMM with ellipsoid
wave-vector surface; (b) The QIMM with double-sheeted wave-vector
surface.}
\end{figure}

In the principal coordinate system, Maxwell's equations yield a
scalar wave equation. In free space, the accompanying dispersion
relation has the familiar form
\begin{equation}
 k_{x}^2+ k_{y}^2+k_{z}^2= \frac{\omega^2}{c^2}, \label{D1}
\end{equation}
where $k_j$ is the $j$ component of the incident wave vector.

We note that the Maxwell's equations are symmetrical in electric
and magnetic fields. So as far as Maxwell's equations is
concerned, what we can do for electricity we can also do for
magnetism. To achieve the polarization insensitive effect, we will
focus our interesting on the anisotropic metamaterial, in which
the permittivity and permeability tensor elements satisfy the
condition:
\begin{equation}
\frac{\varepsilon_x
(\omega)}{\mu_x(\omega)}=\frac{\varepsilon_y(\omega)
}{\mu_y(\omega)}=\frac{\varepsilon_z
(\omega)}{\mu_z(\omega)}=C~~~(C>0), \label{QC}
\end{equation}
where $C$ is a constant. A careful calculation of the Maxwell's
equations gives the dispersion relation:
\begin{widetext}
\begin{equation}
\left(\frac{ q_{x}^2}{\varepsilon_y
\mu_z}+\frac{q_{y}^2}{\varepsilon_x
\mu_z}+\frac{q_{z}^2}{\varepsilon_y \mu_x}-
\frac{\omega^2}{c^2}\right)\left(\frac{ q_{x}^2}{\varepsilon_z
\mu_y}+\frac{q_{y}^2}{\varepsilon_z
\mu_x}+\frac{q_{z}^2}{\varepsilon_x \mu_y}-
\frac{\omega^2}{c^2}\right)=0, \label{D2}
\end{equation}
\end{widetext}
where $q_{j}$ represents the $j$ component of transmitted
wave-vector. The above equation can be represented by a
three-dimensional surface in wave-vector space. This surface is
known as the normal surface and consists of two
shells~\cite{Chen1983}. Under the condition of Eq.~(\ref{QC}), we
can find E- and H-polarized waves exhibit the same wave-vector
surface. Thus the anisotropic medium also be regard as
QIMM~\cite{Shen2005,Luo2006b}. Clearly, we can find the dispersion
surface has the following two types: ellipsoid and double-sheeted
hyperboloid, as show in Fig.~\ref{Fig1}.

We are currently investigating the possibilities for the
manufacture of the QIMM. In fact, it is now conceivable that a
metamaterial can be constructed whose permittivity and
permeability values may be designed to vary independently and
arbitrarily throughout a metamaterial, taking positive or negative
values as desired. Hence the permittivity and permeability tensor
elements of QIMM can be controlled by modulating the length scale
of the SRRs. The permittivity $\varepsilon_j(\omega)$ and
permeability $\mu_j(\omega)$ can be approximated by the Lorentz
model. We trust that the QIMM can be constructed, because similar
technology has been exploited in anisotropic
metamaterial~\cite{Parazzoli2003,Zhou2003,Thomas2005,Grzegorczyk2006}.
In addition, photonic crystals might be a good candidate for
constructing a QIMM. The periodicity in photonic crystals is on
the order of the wavelength, so that the distinction between
refraction and diffraction is blurred. Nevertheless, many novel
dispersion relationships can be realized in photonic crystals,
including ranges where the frequency disperses negatively with
wave vector as required for a negative
refraction~\cite{Notomi2000,Luo2002a,Luo2002b,Li2003}.

Now we want to enquire: whether E- and H-polarized exhibit the
same propagation characteristic. To answer the question we first
discuss the transmission of wave vector. We choose the $z$ axis to
be normal to the interface, the $x$ and $y$ axes locate at in the
plane of the interface. The $z$-component of the transmitted wave
vector can be found by the solution of Eq.~(\ref{D2}), which
yields
\begin{equation}
{q_z^E} = \sigma\sqrt {\varepsilon_y \mu_x k_0^2-\varepsilon_y
\mu_x \left(\frac{q_{x}^2}{\varepsilon_y
\mu_z}+\frac{q_{y}^2}{\varepsilon_x \mu_z}\right)}, \label{qze}
\end{equation}
\begin{equation}
 q_z^H = \sigma\sqrt {\varepsilon_x \mu_y k_0^2-\varepsilon_x \mu_y
\left(\frac{q_{x}^2}{\varepsilon_z\mu_y}
+\frac{q_{y}^2}{\varepsilon_z\mu_x}\right)}, \label{qzh}
\end{equation}
for E- and H-polarized waves, respectively. Here $k_0=\omega/c$
is the wave number in vacuum and $\sigma=\pm 1$.
This choice of sign ensures that power propagates away from the
boundary to the $+z$ direction.

Without loss of generality, we assume the wave vector locate at
the $x-z$  plane ($k_y=q_y=0$). The incident angle of light is
given by
\begin{equation}
\theta_I =\tan^{-1}\left[\frac{k_x}{k_{z}}\right]. \label{IA}
\end{equation}
The values of refractive wave vector can be found by using
boundary conditions and dispersion relations. The refractive angle
of the transmitted wave vector or phase of E- and H- polarized
waves can be written as
\begin{equation}
\beta_P^E=\tan^{-1}\left[\frac{q_x^E}{q_z^E}\right],
~~~\beta_P^H=\tan^{-1}\left[\frac{q_x^H}{q_z^H}\right].\label{AP}
\end{equation}
Substituting Eqs.~(\ref{qze}) and (\ref{qzh}) into Eq.~(\ref{AP}),
we can easily find that E- and H- polarized waves propagates with
the same wave vector or phase velocity.

In the next step, let us discuss the transmission of  energy flux.
It should be noted the actual direction of light is determined by
the time-averaged Poynting vector ${\bf S} =\frac{1}{2} {\bf
Re}({\bf E}^\ast\times \bf{H})$. For E- and H-polarized waves, the
transmitted Poynting vector ${\bf S}_T$ is given by
\begin{equation}
{\bf S}_T^E=Re \left[\frac{T_E^2 E_0^2 q_x^E}{2 \omega \mu_z}{\bf
e}_x+\frac{T_E^2 E_0^2 q_z^E}{2\omega\mu_x}{\bf
e}_z\right],\label{SE}
\end{equation}
\begin{equation}
{\bf S}_T^H=Re \left[\frac{T_H^2 E_0^2 q_x^H}{2 \omega
\varepsilon_z}{\bf e}_x+\frac{T_H^2 E_0^2
q_z^H}{2\omega\varepsilon_x}{\bf e}_z\right],\label{SH}
\end{equation}
where $T_E$ and $T_H$ are the transmission coefficients for E- and
H-polarized waves, respectively. The refraction angle of Poynting
vector of E- and H- polarized incident waves can be obtained as
\begin{equation}
\beta_S^E=
\tan^{-1}\left[\frac{S_{Tx}^E}{S_{Tz}^E}\right],~~~\beta_S^H=
\tan^{-1}\left[\frac{S_{Tx}^H}{S_{Tz}^H}\right].\label{AS}
\end{equation}
Combining Eqs.~(\ref{AP}) and (\ref{AS}) we can easily find that
E- and H- polarized waves have the same Poynting vector. As for
the QIMM slab, the refraction at the second interface can be
investigated by the similar procedures.

\begin{figure}
\includegraphics[width=8cm]{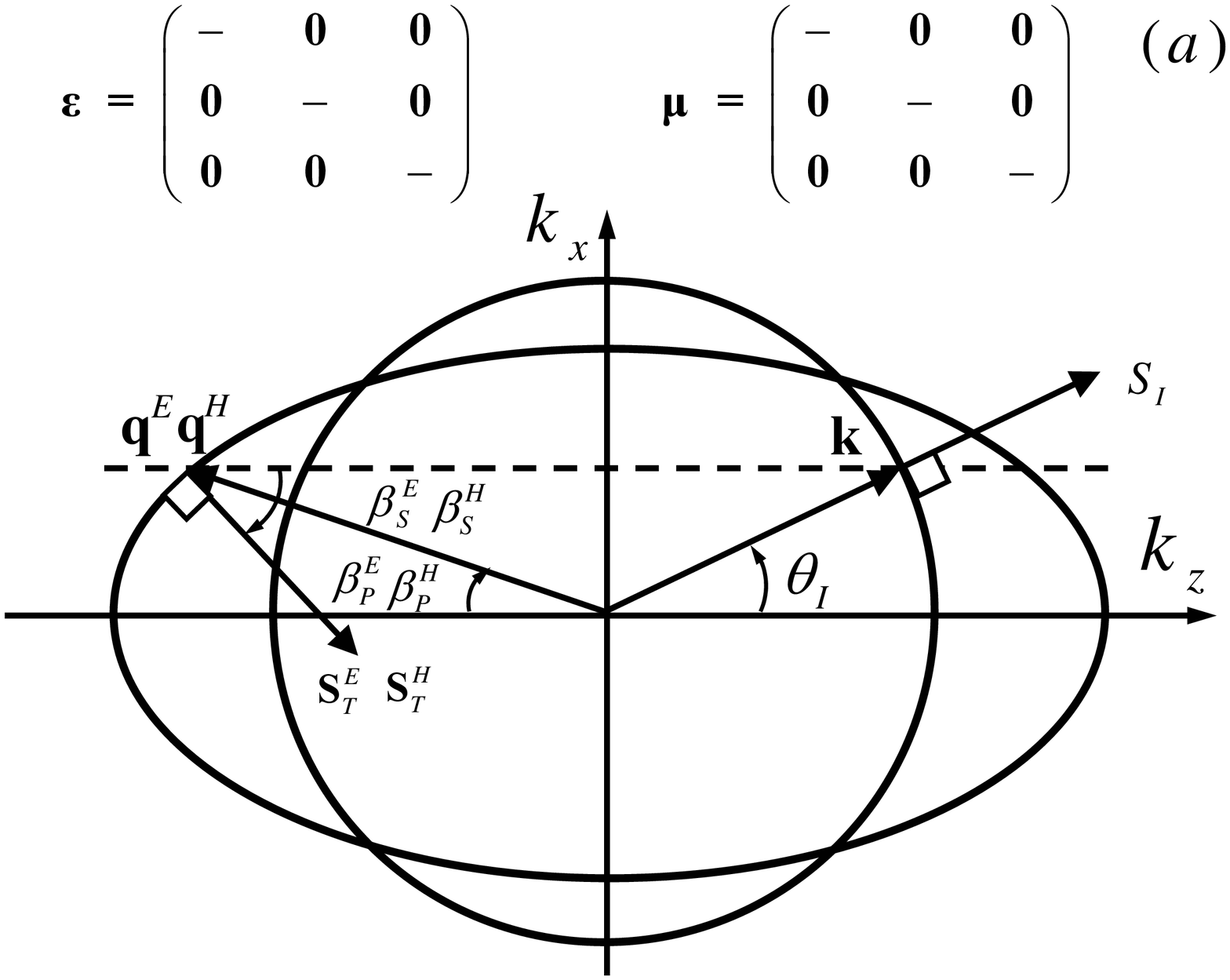}~~~~~
\includegraphics[width=8cm]{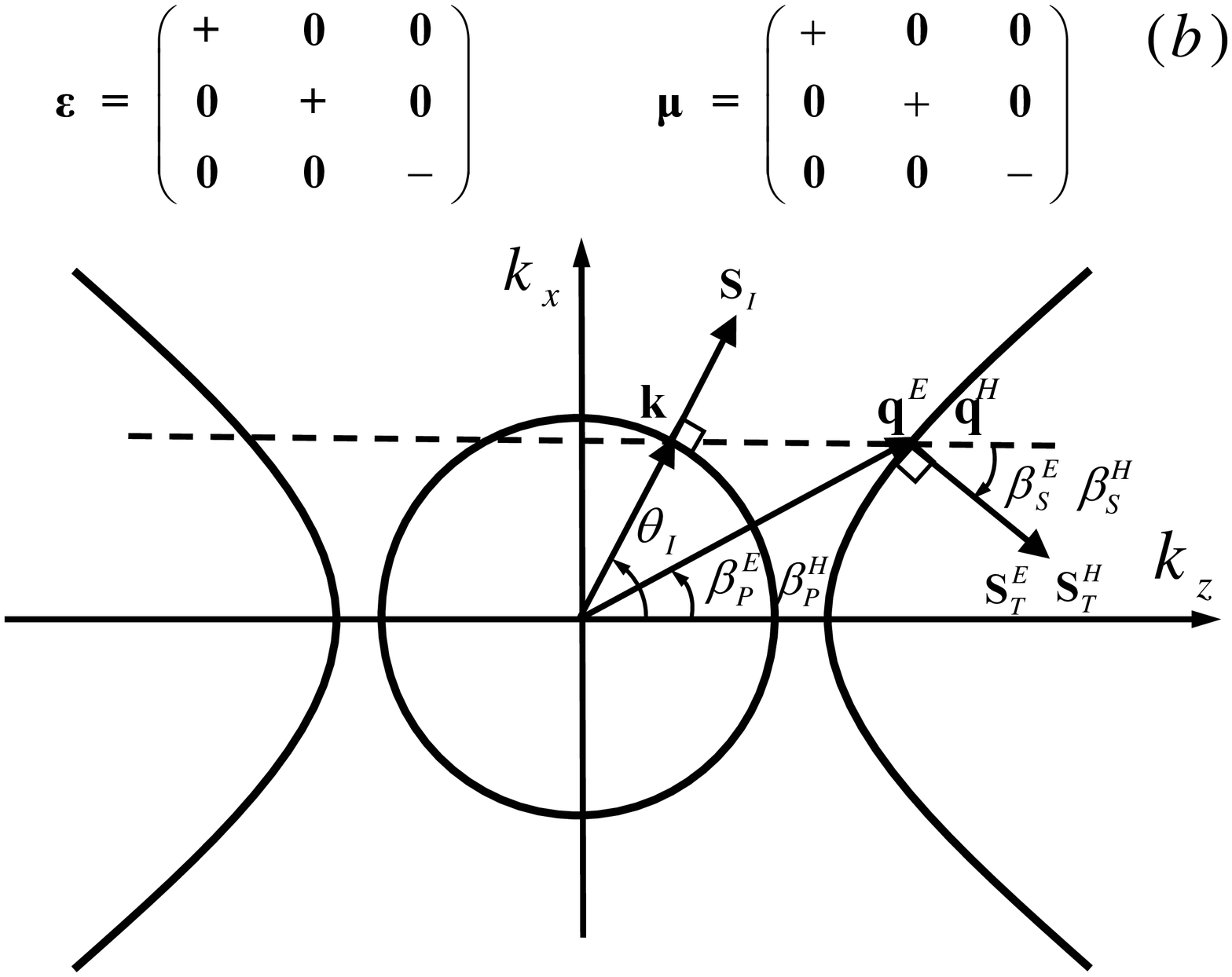}
% Here is how to import EPS art
\caption{\label{Fig2} The frequency contours of isotropic and
quasiisotropic media: (a) The circle and ellipse represent the
frequency contours of vacuum and quasiisotropic media,
respectively. Both the wave vector and the energy flow exhibit
negative refraction. (b) The circle and hyperbola denote the
frequency contours of vacuum and quasiisotropic media,
respectively. The wave vector undergoes a positive refraction,
while the energy flow undergoes a negative refraction.}
\end{figure}

By now, we know that E- and H- polarized waves propagate with same
wave vector and Poynting vector. It is a significantly different
property from general anisotropic media. Note that there is a
bending angle between ${\bf q}$ and ${\bf S}$, and therefore ${\bf
q}$, ${\bf E}$ and ${\bf H}$ do not form a strictly right-handed
or left-handed system in QIMM. Hence it is also different from
isotropic media. For this reason, this kind of special anisotropic
media is regarded as quasiisotropic. It should be mentioned that
if $C<0$ in Eq.~(\ref{QC}), E- and H-polarized waves will exhibit
the same single-sheeted dispersion relation. While the two
polarized waves will undergo different amphoteric refraction, the
special anisotropic media cannot be regarded as
quasiisotropic~\cite{Luo2006b}.

Now we are in the position to study the negative refraction in the
QIMM. Unlike in isotropic media, the Poynting vector in The QIMM
is neither parallel nor antiparallel to the wave vector, but
rather makes either an acute or an obtuse angle with respect to
the wave vector. In general, to distinguish the positive and
negative refraction in QIMM, we must calculate the direction of
the Poynting vector with respect to the wave vector. Positive
refraction means ${\bf q}_x\cdot{\bf S}_{T}>0$, and negative
refraction means ${\bf q}_x\cdot{\bf S}_{T}<0$. From
Eqs.~(\ref{SE}) and (\ref{SH}) we get
\begin{equation}
{\bf q}_x^E\cdot{\bf S}_{T}^{E}=\frac{T_E^2 E_0^2 q_x^2}{2 \omega
\mu_z},~~~{\bf q}_x^H\cdot{\bf S}_{T}^{H}=\frac{T_H^2 H_0^2
q_x^2}{2 \omega \varepsilon_z}.
\end{equation}
The negative refraction phenomenon is one of the most interesting
properties of the QIMM. We can see that the refracted waves will
be determined by $\mu_z$ for E-polarized incident waves and
$\varepsilon_z$  for H-polarized incident waves.

Because of the importance of negative refraction in refocusing
effect, we are interested in the two types of QIMM, which can
formed from appropriate combinations of material parameter tensor elements. \\
Type I.~ In this case all of the $\varepsilon_j$ and $\mu_j$ are
negative. The frequency contour is an ellipse as shown in
Fig.~\ref{Fig2}(a).  Here ${\bf k}_z\cdot{\bf q}_{z}<0$ and ${\bf
q}_x\cdot{\bf S}_T<0$, so the
refraction angle of wave vector and Poynting vector are always negative. \\
Type II. In this case $\varepsilon_x>0$, $\varepsilon_y>0$ and
$\varepsilon_z<0$.  The frequency contour is a double-sheeted
hyperbola as depicted in Fig.~\ref{Fig2}(b).  Here ${\bf
k}_z\cdot{\bf q}_{z}>0$ and ${\bf q}_x\cdot{\bf S}_{T}<0$. It
yields that the refraction of Poynting vector refraction is always
negative even if the wave-vector refraction is positive.

As noted above, the Poynting vector will exhibit negative
refraction in the two types of QIMM. The negative refraction is
the important effect responsible for the slab lens. Hence, the two
kind of QIMM can be employed to construct a polarized insensitive
lens.

\section{The Paraxial model of beam propagation }\label{II}
\begin{figure}
\includegraphics[width=10cm]{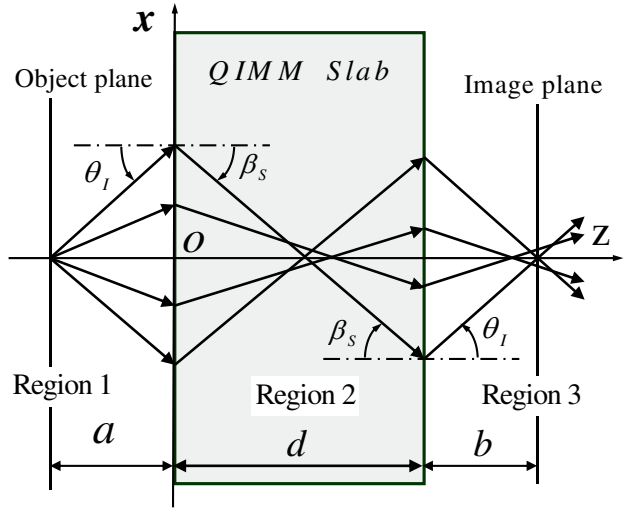}% Here is how to import EPS art
\caption{\label{Fig3}  The ray tracing picture showing the
focussing by QIMM slab. The QIMM slab is surrounded by vacuum in
region 1 and region 3. The solid line and dash-dotted lines are
the theoretical objective and focusing planes, respectively.}
\end{figure}

In this section, we consider the slab lens constructed by the
QIMM. As depicted in Fig.~\ref{Fig3}, the QIMM slab in region $2$
is surrounded by vacuum in region $1$ and region $3$. A point
source is place the object plane $z=0$. A single ray will pass the
interfaces $z = a$ and $z = a+d$ before it reaches the image plane
$z=a+b+d$. Let us investigate what happens when the single ray
pass through the QIMM slab. Because of the anisotropy the ray
incident for different angle will exhibit different image plane.

First, we explore the aberration effect caused by the anisotropic
effect. Subsequent calculations of Eq.~(\ref{AS}) give the
relationship between $\sin{\theta_I}$ and $\sin \beta_s$ by
\begin{equation}
\sin^2 \beta_S^E= \frac{\mu_x^2 \sin^2\theta_I}{\varepsilon_y
\mu_z^2+(\mu_x-\mu_z)\sin^2\theta_I},\label{DESA}
\end{equation}
\begin{equation}
\sin^2 \beta_S^H= \frac{\varepsilon_x^2 \sin^2\theta_I}{\mu_y
\varepsilon_z^2+(\varepsilon_x-\varepsilon_z)\sin^2\theta_I}.\label{DHSA}
\end{equation}
Note that the expressions are sightly different from those in
conventional uniaxial crystal~\cite{Dumelow2005}. It is
instructive to compare these results with Snell's law, which for
refraction from vacuum into an isotropic medium, gives $\sin
\theta_I / \sin\beta_s=n$, where $n$ represents the refractive
index of the refracting medium. Evidently, in the present case we
can find the relationship between $\sin{\theta_I}$ and $\sin
\beta_s$ is nonlinear, which is caused by the anisotropic effect
$\varepsilon_x\neq\varepsilon_z$ or $\mu_x\neq\mu_z$. The
nonlinear relationship will result in a significant aberration
effect in the image plane, hence the achievable resolution of
image is limited.

\begin{figure}
\includegraphics[width=10cm]{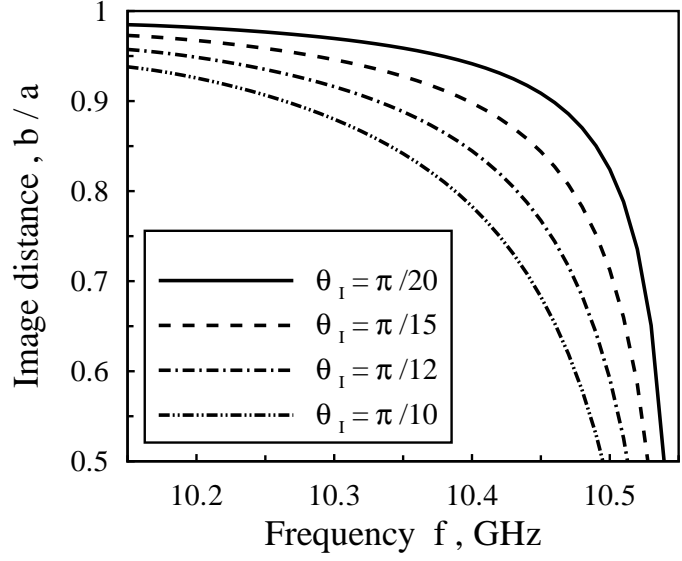}% Here is how to import EPS art
\caption{\label{Fig4} For a certain incident angle, the ray with
different frequency $f=\omega/2\pi$ will exhibit different image
plane. Note that for smaller incident angles, the QIMM slab will
provide some degree of frequency insensitive in a certain GHz
band.}
\end{figure}

Next, we want to discuss another aberration effect caused by
frequency dispersion. For a certain incident angle, the ray with
different frequency will exhibit different image plane as shown in
Fig.~\ref{Fig4}. We assume that both $\varepsilon_z(\omega)$ and
$\mu_z(\omega)$ be approximated by the Lorentz model. The other
tensor elements components are approximated as constants. The
resonate frequency, plasma frequency, and damping constant are
identical in all respects to those utilized in
Ref.~\cite{Thomas2005}. Ignoring the metallic structure, the other
tensor elements assume the values of the background material which
is dominantly air. Obviously the frequency dispersion of
$\varepsilon_z(\omega)$ and $\mu_z(\omega)$ will place some
practical limitations on the resolution of image. Fortunately the
limitations can be reduced, if the ray incident at a small angle.
The significant effect has be illustrated in Fig.~\ref{Fig4}. It
is interesting to noted that the image distance will slowly vary
with frequency in a certain band.

Remarkable as the QIMM slab lens is it suffers from some problem:
how to cancel the aberration effect?  Does the polarization
insensitive lens have any use, if the image is not imperfect? The
main effect of anisotropy associated with QIMM will limit the
resolution of the image. Although the image is imperfect, we trust
that the reconstruct effect of intensity and phase in paraxial
regime will lead to some potential applications. Such as the QIMM
slab can be used to provide phase compensation and beam focusing
in cavity resonator. Furthermore, sightly anisotropy in QIMM slab
lens can improve the beam parameter. Most importantly, the QIMM
can be used to design polarization-insensitive modulators and
polarization-insensitive all-optical switching in fiber
communication system. Hence it is very desirable to investigate
the polarization insensitive effect in paraxial regime.

From a mathematical point of view, the approximate paraxial
expression for the field can be obtained by the expansion of the
square root of $q_z$ to the first order in $|{\bf
q}_\bot|/q$~\cite{Lax1975,Ciattoni2000,Luo2006c}, which yields
\begin{equation}
 q_z^E = \sigma \sqrt{\varepsilon_y \mu_x} k_0-\left(\frac{\sigma  \sqrt{\varepsilon_y \mu_x} k_x^2}
{2 \varepsilon_y \mu_z k_0}+\frac{\sigma \sqrt{\varepsilon_y
\mu_x} k_y^2}{2 \varepsilon_x \mu_z k_0}\right), \label{pqze}
\end{equation}
\begin{equation}
 q_z^H = \sigma \sqrt{\varepsilon_x \mu_y} k_0-\left(\frac{\sigma  \sqrt{\varepsilon_x \mu_y} k_x^2}{2
\varepsilon_z\mu_y k_0}+\frac{\sigma \sqrt{\varepsilon_x \mu_y}
k_y^2}{2 \varepsilon_z \mu_x k_0}\right), \label{pqzh}
\end{equation}
where we have introduced the boundary condition ${\bf q}_\bot={\bf
k}_\bot$. From Eqs.~(\ref{pqze}) and (\ref{pqzh}) we can easily
find that for shallow incident angles the QIMM slab will provide
some degree of refocusing in the same manner as an isotropic LHM.
Hence the aberration effect can be cancelled in paraxial beam
region. The interesting property allow us to introduce the idea to
construct a QIMM slab lens in paraxial beam region.

Equation~(\ref{mpe}) can be conveniently solved by employing the
Fourier transformations, so the complex amplitudes in QIMM for E-
and H-polarized beams can be conveniently expressed as
\begin{equation}
 {\bf E}({ {\bf r}_\bot},z )=\int d^2 {\bf k}_\bot
\tilde{E}({\bf k}_{\bot})\exp [i{\bf k}_{\bot}\cdot{\bf
r}_{\perp}+i q_z^E z].\label{fielde}
\end{equation}
\begin{equation}
 {\bf H}({ {\bf r}_\bot},z )=\int d^2 {\bf k}_\bot
\tilde{H}({\bf k}_{\bot})\exp [i{\bf k}_{\bot}\cdot{\bf
r}_{\perp}+i q_z^H z].\label{fieldh}
\end{equation}
Here ${\bf r}_\perp=x{\bf e}_x+y{\bf e}_y$, ${\bf k}_\perp
=k_x{\bf e}_x+k_y {\bf e}_y$, and $e_j$ is the unit vector in the
$j$-direction.

Substituting Eqs.~(\ref{pqze}) and (\ref{pqzh}) into
Eq.~(\ref{fielde}) and (\ref{fieldh}), respectively, we obtain
\begin{widetext}
\begin{eqnarray}
 {\bf E}({ {\bf r}_\bot},z )=&&\exp(i
\sigma \sqrt{\varepsilon_x \mu_y} k_0 z) \int d^2 {\bf k}_\bot\nonumber\\
&&\times\exp \bigg[i{\bf k}_{\perp}\cdot{\bf
r}_{\perp}-\left(\frac{\sigma \sqrt{\varepsilon_y \mu_x}}{2
\varepsilon_y \mu_z k_0} k_x^2+ \frac{\sigma \sqrt{\varepsilon_y
\mu_x}}{2 \varepsilon_x \mu_z k_0} k_y^2\right)\bigg]\tilde{\bf
E}({\bf k}_\perp),\label{fieldE}
\end{eqnarray}
\begin{eqnarray}
 {\bf H}({ {\bf r}_\bot},z )=&&\exp(i
\sigma \sqrt{\varepsilon_x \mu_y} k_0 z) \int d^2 {\bf k}_\bot\nonumber\\
&&\times\exp \bigg[i{\bf k}_{\perp}\cdot{\bf
r}_{\perp}-\left(\frac{\sigma \sqrt{\varepsilon_y \mu_x}}{2
\varepsilon_z \mu_y k_0} k_x^2+ \frac{\sigma \sqrt{\varepsilon_y
\mu_x}}{2 \varepsilon_z \mu_x k_0} k_y^2\right)\bigg]\tilde{\bf
H}({\bf k}_\perp).\label{fieldH}
\end{eqnarray}
\end{widetext}
The fields $\tilde{E}({\bf k}_{\bot})$ and $\tilde{H}({\bf
k}_{\bot})$  in Eqs.~(\ref{fieldE}) and (\ref{fieldH}) are related
to the boundary distributions of the fields by means of the
relation
\begin{equation}
\tilde{{\bf E}}({\bf k}_{\bot})=\int d^2 {\bf r}_\bot {\bf
E}({\bf r}_{\bot},0)\exp [i{\bf k}_{\bot}\cdot{\bf r}_{\perp}]
,\label{asE}
\end{equation}
\begin{equation}
\tilde{{\bf H}}({\bf k}_{\bot})=\int d^2 {\bf r}_\bot {\bf
H}({\bf r}_{\bot},0)\exp [i{\bf k}_{\bot}\cdot{\bf r}_{\perp}]
,\label{asH}
\end{equation}
for E- and H-polarized beams, respectively. Evidently,
Eqs.~(\ref{asE}) and (\ref{asH}) are standard two-dimensional
Fourier transform~\cite{Goodman1996}. In fact, after the field
distribution in the plane $z=0$ is known, Eqs.~(\ref{fieldE}) and
(\ref{fieldH}) provide the expression of the E- and H-polarized
field in the space $z > 0$, respectively.

Since our attention will be focused on beam propagating along the
$+z$ direction, we can write the paraxial fields as
\begin{equation}
{\bf E}({\bf r}_\perp,z)={\bf A}_E({\bf r}_\perp,z) \exp(i
\sigma \sqrt{\varepsilon_y \mu_x} k_0 z),\label{peq}
\end{equation}
\begin{equation}
{\bf H}({\bf r}_\perp,z)={\bf A}_H({\bf r}_\perp,z) \exp(i
\sigma \sqrt{\varepsilon_x \mu_y} k_0 z),\label{peq}
\end{equation}
where the field $A( {\bf r}_\perp , z)$ is the slowly varying
envelope amplitude which satisfies the parabolic equation:
\begin{equation}
\left[i\frac{\partial}{\partial z}+\left(\frac{\sigma
\sqrt{\varepsilon_y \mu_x}}{2 \varepsilon_y \mu_z k_0}
\frac{\partial^2}{\partial x^2}+ \frac{\sigma \sqrt{\varepsilon_y
\mu_x}}{2 \varepsilon_x \mu_z k_0} \frac{\partial^2}{\partial
y^2}\right)\right]{\bf
 A}_E({\bf r}_\perp,z)=0,\label{pee}
\end{equation}

\begin{equation}
\left[i\frac{\partial}{\partial z}+\left(\frac{\sigma
\sqrt{\varepsilon_x \mu_y}}{2 \varepsilon_z \mu_y k_0}
\frac{\partial^2}{\partial x^2}+ \frac{\sigma \sqrt{\varepsilon_x
\mu_y}}{2 \varepsilon_z \mu_x k_0} \frac{\partial^2}{\partial
y^2}\right)\right]{\bf
 A}_H({\bf r}_\perp,z)=0.\label{peh}
\end{equation}
Under the quasiisotropic condition of Eq.~(\ref{QC}), we can
easily find that E- and H-polarized paraxial filed exhibit the
same propagating characteristics in paraxial regime. The
interesting properties allow us to introduce the idea to construct
a polarization lens by QIMM slab. For simplify, we introduce the
effective refraction indexes:
\begin{equation}
n_x=\sigma \frac{\varepsilon_y \mu_z} {\sqrt{\varepsilon_y
\mu_x}},~~~~~n_y=\sigma \frac{\varepsilon_x \mu_z}
{\sqrt{\varepsilon_y \mu_x}}.
\end{equation}
From Eqs.~(\ref{pee}) and (\ref{peh}) we can find that the field
of paraxial beams in QIMM can be written in the similar way to
that in regular material, while the sign of the effective
refraction index could be reverse. To simplify the proceeding
analyses, we will focus our attention on the QIMM with ellipsoid
frequency contour.

\section{Beam focusing by polarization insensitive lens}\label{SecIII}
In the previous section we have understood both E- and H-polarized
beams have the same propagation characteristic in QIMM slab. Hence
we do not wish to get involved in the trouble to discuss the
focusing effect of two polarized waves. Instead, we will
investigate the analytical description for E-polarized beam with a
boundary Gaussian distribution. This example allows us to describe
the refocusing features of beam propagation in QIMM slab. To be
uniform throughout the following analysis, we introduce different
coordinate transformations $z_i^\ast (i=1,2,3)$ in the three
regions, respectively. First we want to explore the field in
region $1$. Without any loss of generality, we assume that the
input waist locates at the object plane $z=0$. The fundamental
Gaussian spectrum distribution can be written in the form
\begin{equation}
\tilde{{\bf E}}_1( {\bf k}_\perp)=\frac{w_0
E_0}{\sqrt{2\pi}}\exp\bigg[-\frac{k_\perp^2
w_0^2}{4}\bigg],\label{s1}
\end{equation}
where $w_0$ is the spot size. The Rayleigh lengths give by
$z_R=k_0 w_0^2 /2$. By substituting Eq.~(\ref{s1}) into
Eq.~(\ref{fielde}), the field in the region $1$ can be written as
\begin{equation}
{\bf
 E}_{1}({\bf r}_\perp,z_1^\ast)=\frac{w_0 E_0}
 {\sqrt{w_{1x}w_{1y}}}\exp\left[-\left(\frac{x^2}
 {w_{1x}^2}+\frac{y^2}
 {w_{1y}^2}\right)+i \psi_1\right],\label{g1}
\end{equation}
\begin{eqnarray}
w_{1x}=w_0 \sqrt{1+\left(\frac{z_{1x}^\ast}{L_{1x}}\right)^2},~~~
w_{1y}=w_0\sqrt{1+\left(\frac{z_{1y}^\ast}{L_{1y}}\right)^2}.\label{w1}
\end{eqnarray}
Here we have chosen different waists, $w_{1x}$ and $w_{1y}$, in
order to deal with a more general situation. Because of the
isotropy in vacuum, we can easily obtain
$z_{1x}^\ast=z_{1y}^\ast=z$ and $w_{1x}=w_{1y}$. The corresponding
Rayleigh lengths give by $L_{1x}=L_{1y}=z_R$.

We are now in a position to investigate the field in region $2$.
In fact, the field in the first boundary can be easily obtained
from Eq.~(\ref{g1}) by choosing $z=a$. Substituting the field into
Eq.~(\ref{asE}), the angular spectrum distribution can be obtained
as
\begin{equation}
\tilde{{\bf E}}_{2}( {\bf k}_\perp)=\frac{w_0
E_0}{\sqrt{2\pi}}\exp\left[-\frac{ k_0 w_0^2+2ia}{4
k_0}(k_x^2+k_y^2)\right].\label{s2}
\end{equation}
For simplicity, we assume that the wave propagate through the
boundary without reflection. Substituting Eq.~(\ref{s2}) into
Eq.~(\ref{fieldE}), the field in the QIMM slab can be written as
\begin{equation}
{\bf
 E}_{2}({\bf r}_\perp,z_2^\ast)=\frac{w_0 E_0}
 {\sqrt{w_{2x}w_{2y}}}\exp\left[-\left(\frac{x^2}
 {w_{2x}^2}+\frac{y^2}
 {w_{2y}^2}\right)+i \psi_2\right],\label{g2}
\end{equation}
\begin{eqnarray}
w_{2x}=w_0 \sqrt{1+\left(\frac{z_{2x}^\ast}{L_{2x}}\right)^2},~~~
w_{2y}=w_0\sqrt{1+\left(\frac{z_{2y}^\ast}{L_{2y}}\right)^2}.\label{w2}
\end{eqnarray}
Here $z_{2x}^\ast=z-(1-n_x)a$ and $z_{2y}^\ast=z-(1-n_y)a$. The
interesting point we want to stress is that there are two
different Rayleigh lengths, $L_{2x}=n_x k_0 w_0^2 /2$ and
$L_{2y}=n_y k_0 w_0^2 /2$, that characterize the spreading of the
beam in the direction of $x$ and $y$ axes, respectively. A further
important point should be noted that we have introduce the
negative Rayleigh length. The inherent physics underlying the
negative Rayleigh length is the waves undergo a negative phase
velocity in the QIMM slab. As can be seen in the following, the
negative Rayleigh length will give rise to the corresponding
reverse Gouy phase shift.

\begin{figure}
\includegraphics[width=6cm]{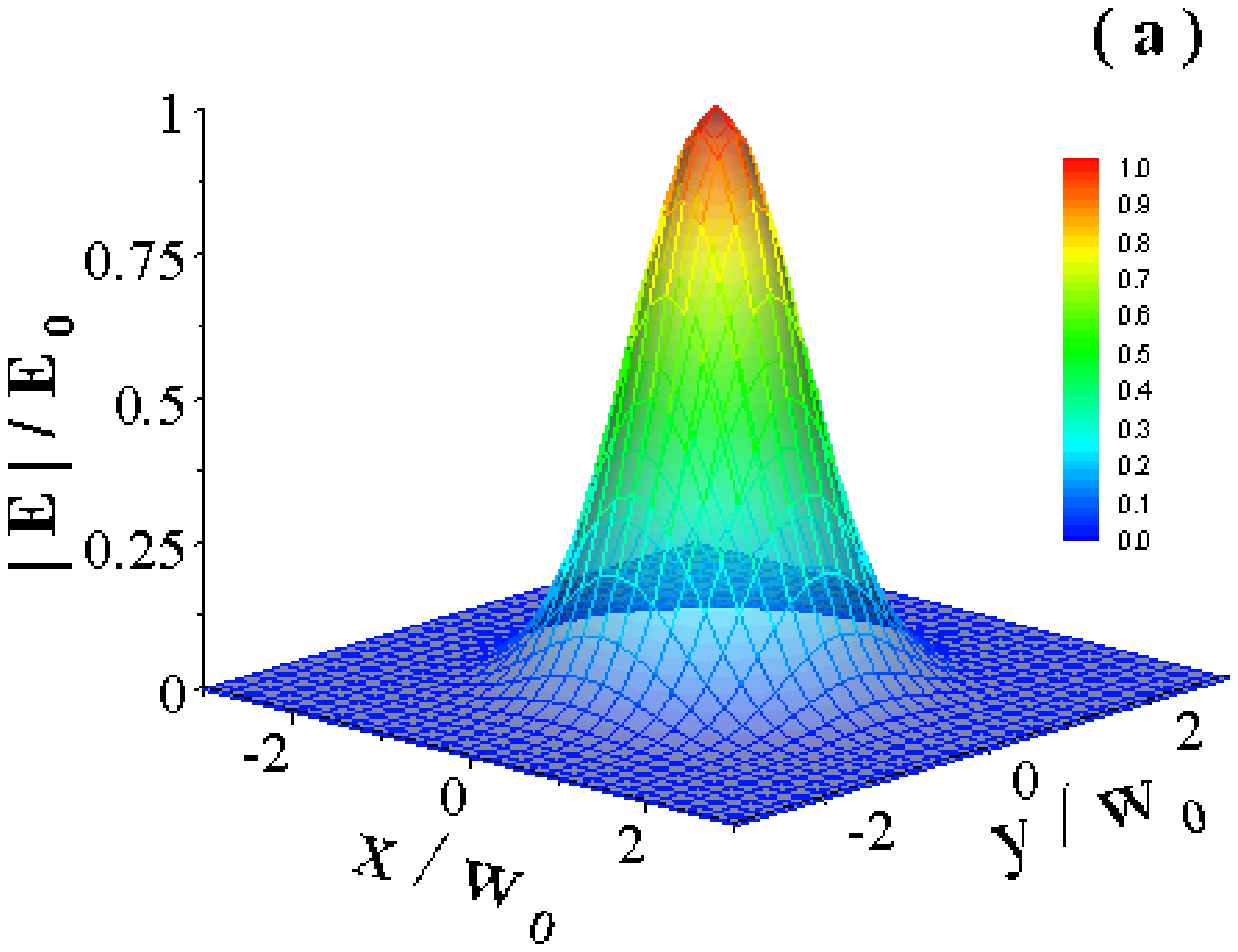}~~~~~
\includegraphics[width=6cm]{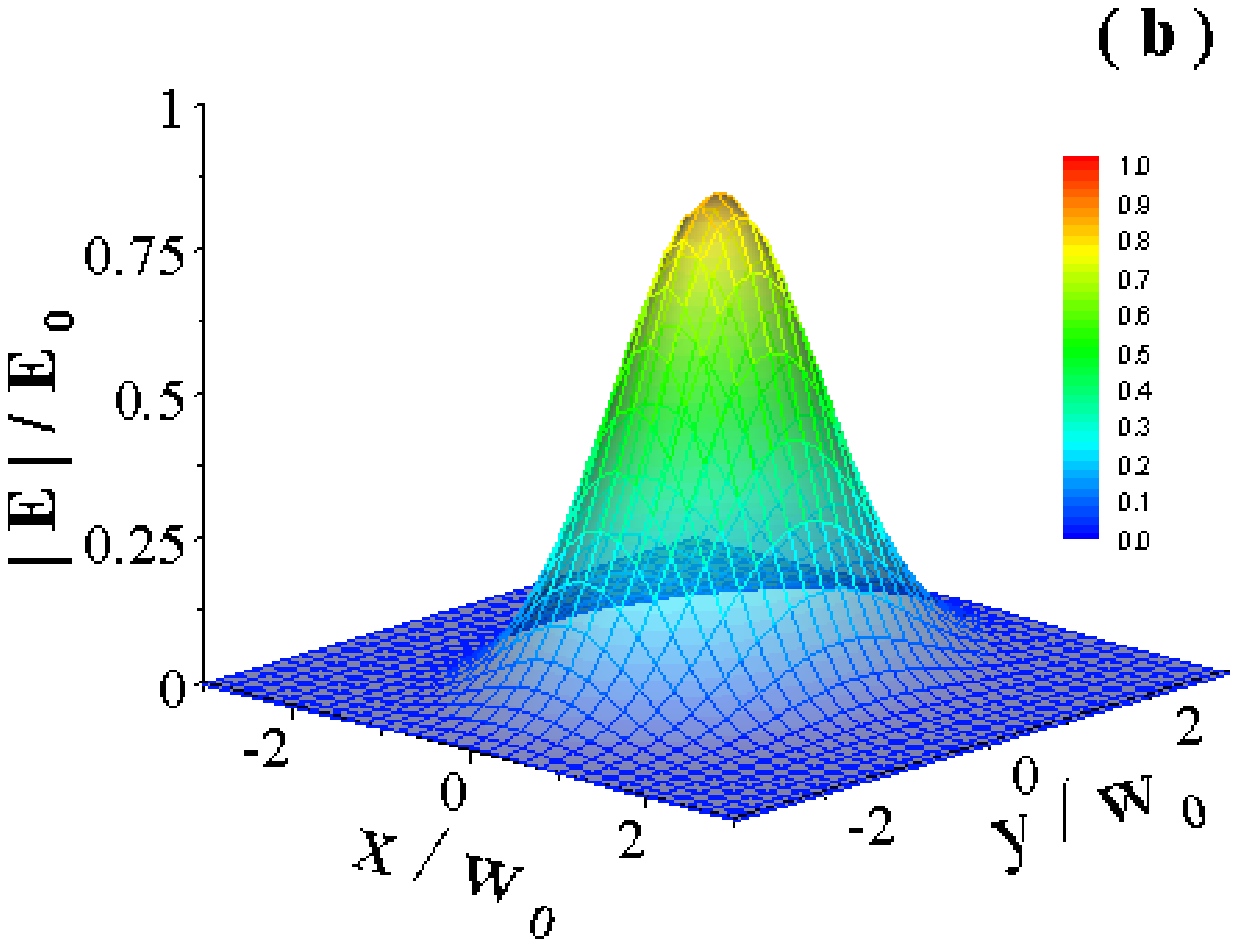}
\includegraphics[width=6cm]{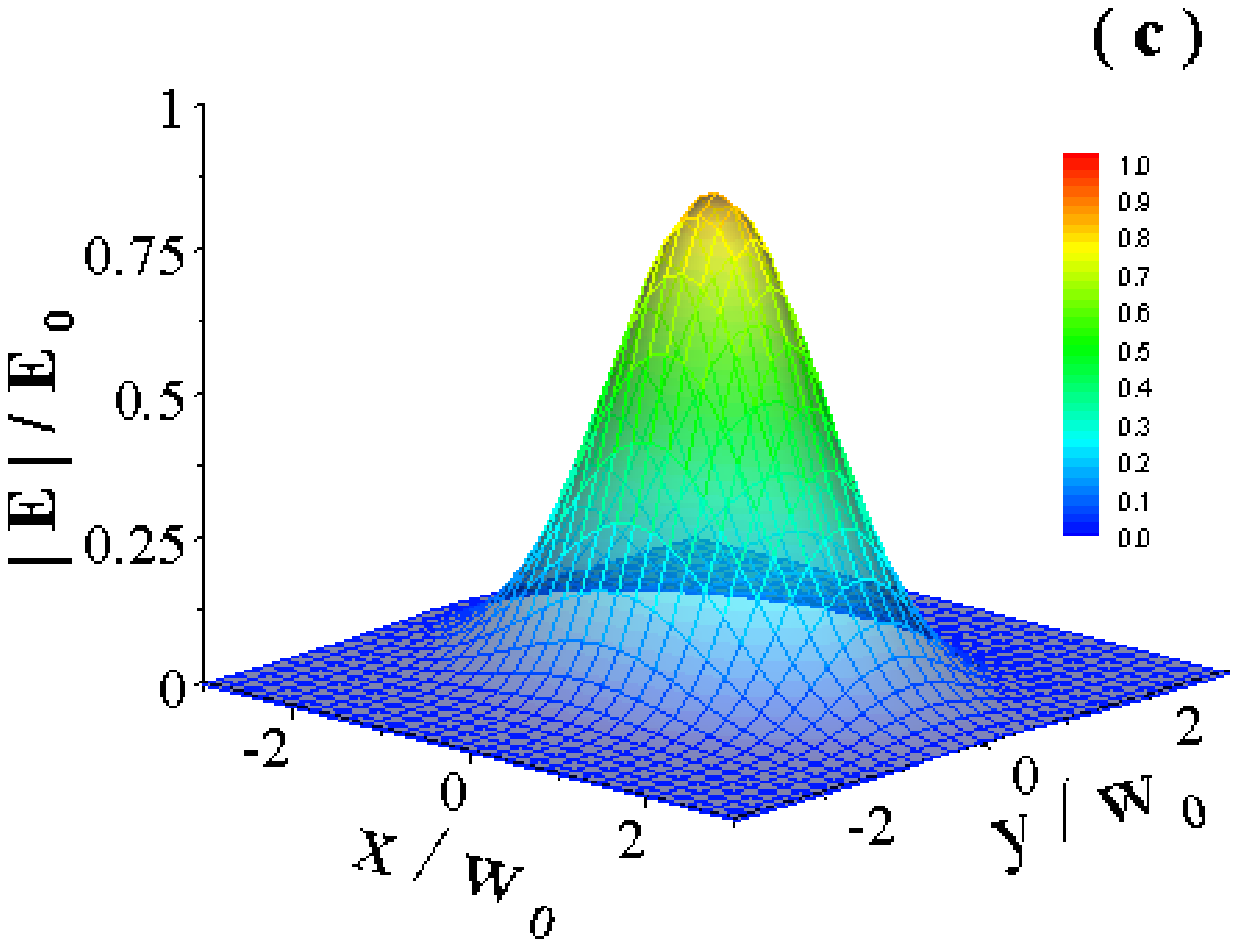}~~~~~
\includegraphics[width=6cm]{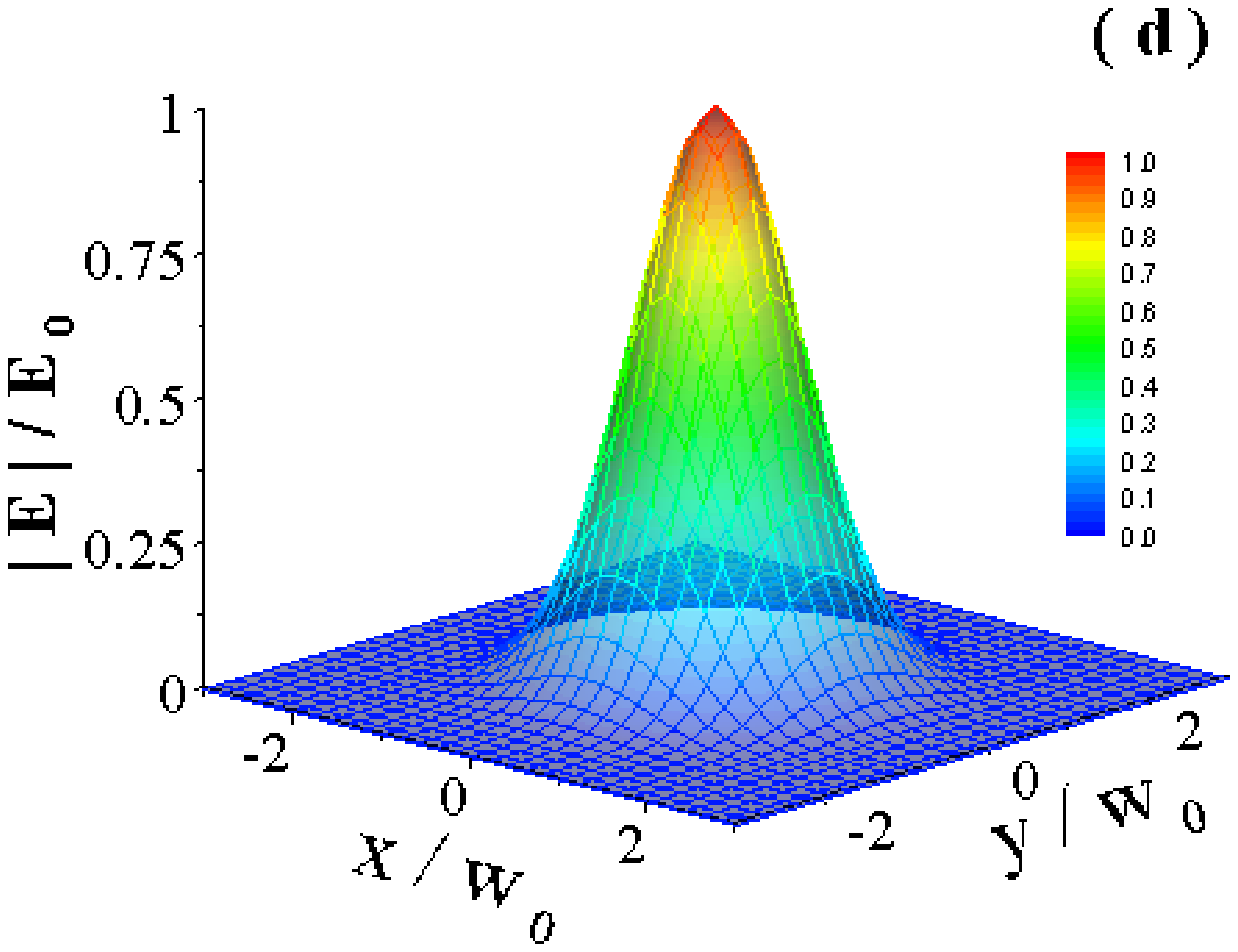}
% Here is how to import EPS art
\caption{\label{Fig5} (Color online) The numerically computed
intensity distribution in object and image planes. (a) the
intensity distribution for normal Gaussian beam in object plane.
The in tensity distribution in image plane for Gaussian beam
propagating through the QIMM slab with different anisotropic
parameters: (b) $n_x=-1$, $n_y=-2$. (c) $n_x=-2$, $n_y=-1$. (d)
$n_x=-1$, $n_y=-1$. We can easily find the intensity distribution
at the object plane can be completely reconstructed at the image
plane.}
\end{figure}

Finally we want to explore the field in region $3$. The field in
the second boundary can be easily obtained from Eq.~(\ref{g2})
under choosing $z=a+d$. Substituting the field into
Eq.~(\ref{asE}), the angular spectrum distribution can be written
as
\begin{widetext}
\begin{equation}
\tilde{\bf
 E}_{3}( {\bf k}_\perp)=\frac{w_0 E_0}{\sqrt{2\pi}}\exp\left[-\left(\frac{
n_x k_0 w_0^2+2in_x a+2i d}{4 n_x k_0}k_x^2+\frac{ n_y k_0
w_0^2+2in_y a+2i d}{4 n_y k_0}k_y^2\right)\right].\label{s3}
\end{equation}
\end{widetext}
Substituting Eq.~(\ref{s3}) into Eq.~(\ref{fieldE}), the field in
the region 3 is given by
\begin{equation}
{\bf E}_{3}({\bf r}_\perp,z_3^\ast)=\frac{w_0 E_0}
{\sqrt{w_{3x}w_{3y}}}\exp\left[-\left(\frac{x^2}
{w_{3x}^2}+\frac{y^2} {w_{3y}^2}\right)+i \psi_3\right],\label{g3}
\end{equation}
\begin{eqnarray}
w_{3x}=w_0 \sqrt{1+\left(\frac{z_{3x}^\ast}{L_{3x}}\right)^2},~~~
w_{3y}=w_0\sqrt{1+\left(\frac{z_{3y}^\ast}{L_{3y}}\right)^2}.\label{w3}
\end{eqnarray}
Here $z_{3x}^\ast=z-(1- 1/n_x)d$ and $z_{3y}^\ast=z-(1- 1/n_y)d$.
The corresponding Rayleigh lengths give by $L_{3x}=L_{3y}=k_0
w_0^2 /2$, that denote the beam exhibit the same diffraction
distance in the direction of $x$ and $y$ axes.
The effect of the anisotropic diffraction is that these
two beam widths keep their difference even if the Rayleigh lengths,
$L_{3x}$ and $L_{3y}$, are equal, implying that generally the Gaussian
beam is astigmatic.

Up to now, the fields are determined explicitly in the three
regions. Comparison of Eq.~(\ref{g2}), Eq.~(\ref{g3}) with
Eq.~(\ref{g1}) shows that the field distributions in region $2$
and region $3$ may no longer remain Gaussian. We take the image
position $z=a+d+b$ to be the place of the second focusing waist.
For the purpose of illustration, the intensity distribution in
object plane is plotted in Fig.~\ref{Fig5}(a). In general, the
shape of intensity distribution is distorted in image plane as
shown in Fig.~\ref{Fig5}(b) and Fig.~\ref{Fig5}(c). Careful
evaluation of Eq.~(\ref{g3}) reveals that the secret underlying
the intensity distortion is the anisotropic diffraction.

Now, the most obvious question is whether the intensity
distribution at the object plane can be completely reconstructed
at the image plane. In the next step, we want to explore the
matching condition of focusing. We can easily obtain the place of
the focusing waist by choosing $z^\ast_i=0$. Let us assume the
incident beam waist locates at plane $z=0$. To eliminate the
astigmatic effect, the beam waists should locate at the same
place, namely $z_{3x}^\ast=z_{3y}^\ast$. Using these criterions,
the matching condition for focusing can be written as
\begin{equation}
\varepsilon_y \mu_z (a+b)+ \sigma \sqrt{\varepsilon_x
\mu_y}d=0,~~~\varepsilon_x=\varepsilon_y.\label{foc}
\end{equation}
Under the focusing matching condition, the intensity distribution
at the object plane can be completely reconstructed at the image
plane as shown in Fig.~\ref{Fig5}(d). A further point should be
noted is that the thickness of the QIMM slab should satisfy the
relation $d>\sigma \varepsilon_y \mu_z a/\sqrt{\varepsilon_x
\mu_y}  $, otherwise there is neither an internal nor an external
focus.

\section{Phase compensation by polarization insensitive lens}\label{SecIV}
In this section, we attempt to investigate the matching condition
for phase compensation. In isotropic LHM, plane waves can
propagate with negative phase velocity directed opposite to the
direction of Poynting vector. Hence the phase difference can be
compensated by the LHM
slab~\cite{Veselago1968,Pendry2000,Luo2006b}. However, the
negative tensor parameters associated with QIMM provides a wealth
of opportunities for observing and exploiting negative
phase-velocity behavior.

First let us investigate the phase distribution in region 1. A
more rigorous calculation of Eq.~(\ref{g1}) gives
\begin{eqnarray}
\psi_{1} &=& k_0 z+ \left( \frac{k_0 x^2}{2 R_{1x}}+\frac{k_0
y^2}{2 R_{1y}}\right)-\Phi_1,\label{p1}
\end{eqnarray}
\begin{eqnarray}
R_{1x}&=&z_{1x}^\ast+\frac{L_{1x}^2}{z_{1x}^\ast},~~~R_{1y}=z_{1y}^\ast+\frac{L_{1y}^2}{z_{1y}^\ast},\\
\Phi_1&=&-\frac{1}{2}\left(\arctan
\frac{z_{1x}^\ast}{L_{1x}}+\arctan
\frac{z_{1y}^\ast}{L_{1y}}\right).
\end{eqnarray}
Here $R_{1x}$ and $R_{1y}$ are the radius of
curvature. Because of the isotropy in vacuum, we can easily find
$R(z_{1x}^\ast)=R(z_{1y}^\ast)$. The Gouy phase shift in vacuum is
given by $\Phi_1$.

Next, we attempt to explore the phase distribution in region 2.
Matching the boundary condition, the phase term in Eq.~(\ref{g2})
can be written as
\begin{eqnarray}
\psi_{2} &=& k_0 a+\sigma \sqrt{\varepsilon_y \mu_x} k_0 (z-a)+
\frac{k_0 x^2}{2 R_{2x}}+\frac{k_0 y^2}{2
R_{2y}}-\Phi_2,\label{p2}
\end{eqnarray}
\begin{eqnarray}
R_{2x}&=&z_{2x}^\ast+\frac{L_{2x}^2}{z_{2x}^\ast},~~~R_{2y}=z_{2y}^\ast+\frac{L_{2y}^2}{z_{2y}^\ast},\\
\Phi_2&=&-\frac{1}{2}\left(\arctan
\frac{z_{2x}^\ast}{L_{2x}}+\arctan
\frac{z_{2y}^\ast}{L_{2y}}\right).\label{gouy2}
\end{eqnarray}
The Gouy phase shift in QIMM is given by Eq.~(\ref{gouy2}). We
should mention that there are two different radius of curvature,
$R_{2x}$ and $R_{2y}$, that characterize the beam undergo
different diffraction effects in the direction of $x$ and $y$
axes, respectively.

\begin{figure}
\includegraphics[width=10cm]{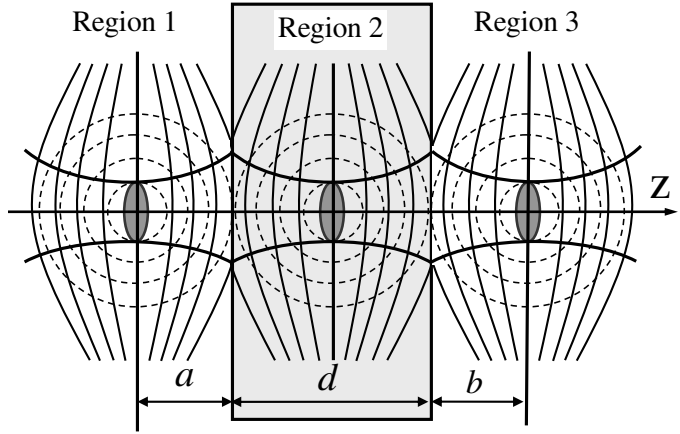}% Here is how to import EPS art
\caption{\label{Fig6} The phase difference caused by the Gouy
phase shift in vacuum can be compensated by that caused by the
inverse Gouy phase shift in the QIMM slab. The phase fronts of
Gaussian beam (solid lines) differ from those of a perfect
spherical wave (dashed lines).}
\end{figure}

Now, we are in the position to explore the phase distribution in
region 3.  Analogously, we make some serious calculation of
Eq.~(\ref{g3}), then obtain the phase distribution
\begin{eqnarray}
\psi_{3} &=& \sigma \sqrt{\varepsilon_y \mu_x} k_0 d+k_0 (z-d)+
\frac{k_0 x^2}{2 R_{3x}}+\frac{k_0 y^2}{2 R_{3y}
}-\Phi_3,\label{p3}
\end{eqnarray}
\begin{eqnarray}
R_{3x}&=&z_{3x}^\ast+\frac{L_{3x}^2}{z_{3x}^\ast},~~~R_{3y}=z_{3y}^\ast+\frac{L_{3y}^2}{z_{3y}^\ast},\label{r3}\\
\Phi_3&=&-\frac{1}{2}\left(\arctan
\frac{z_{3x}^\ast}{L_{3x}}+\arctan
\frac{z_{3y}^\ast}{L_{3y}}\right).\label{gouy3}
\end{eqnarray}
The radius of curvatures are given by Eq.~(\ref{r3}) and the
corresponding Gouy phase shift is given by Eq.~(\ref{gouy3}). The
anisotropic effect result in the two radius of curvatures keep
their difference even if the Rayleigh lengths are equal.

It is known that an electromagnetic beam propagating through a
focus experiences an additional $\pi$ phase shift with respect to
a plane wave. This phase anomaly was discovered by Gouy in 1890
and has since been referred to as the Gouy phase
shift~\cite{Born1997,Feng2001}. It should be mentioned that there exists an
effect of accumulated Gouy phase shift when a beam passing
through an  optical system with positive
index~\cite{Erden1997,Feng1999,Feng2000}. While in the QIMM slab
system we expect that the phase difference caused by Gouy phase
shift can be compensated by that caused by the inverse Gouy shift
in the QIMM slab.

\begin{figure}
\includegraphics[width=6cm]{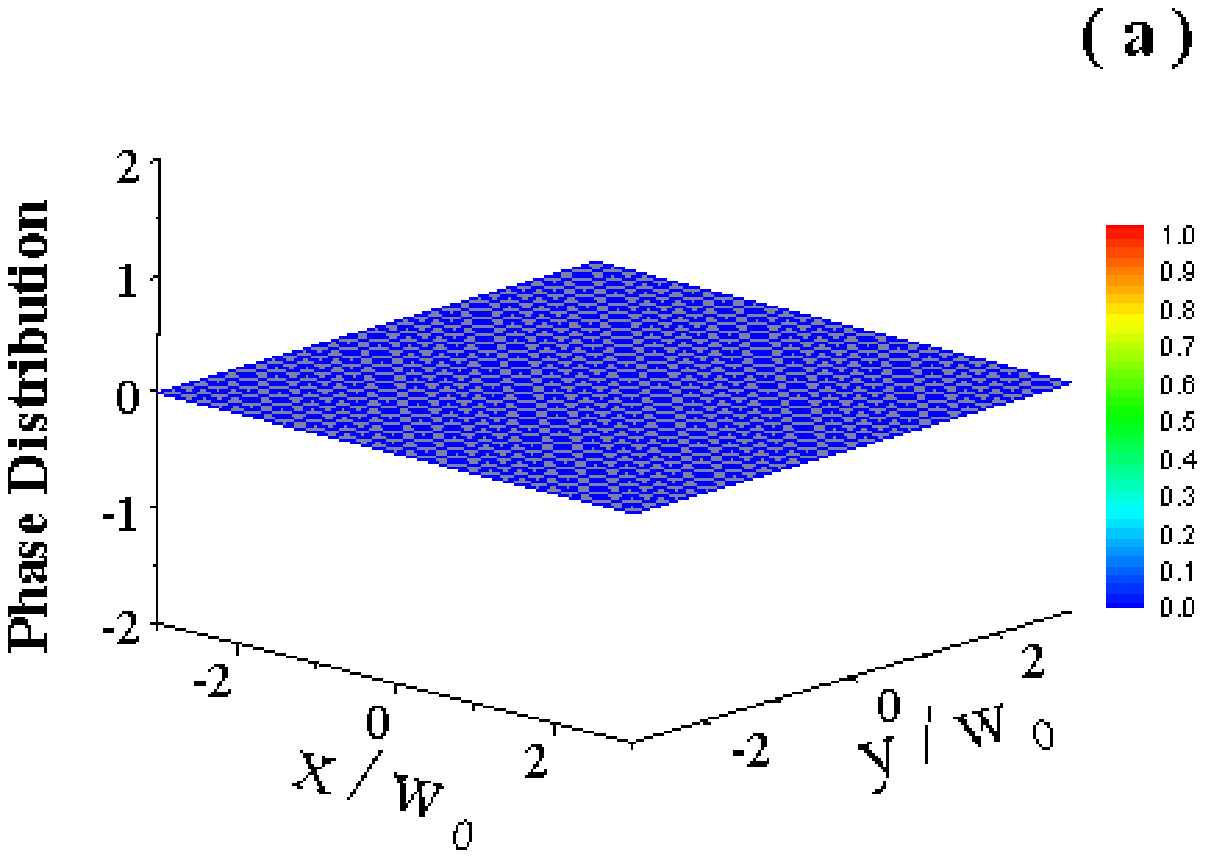}~~~~~
\includegraphics[width=6cm]{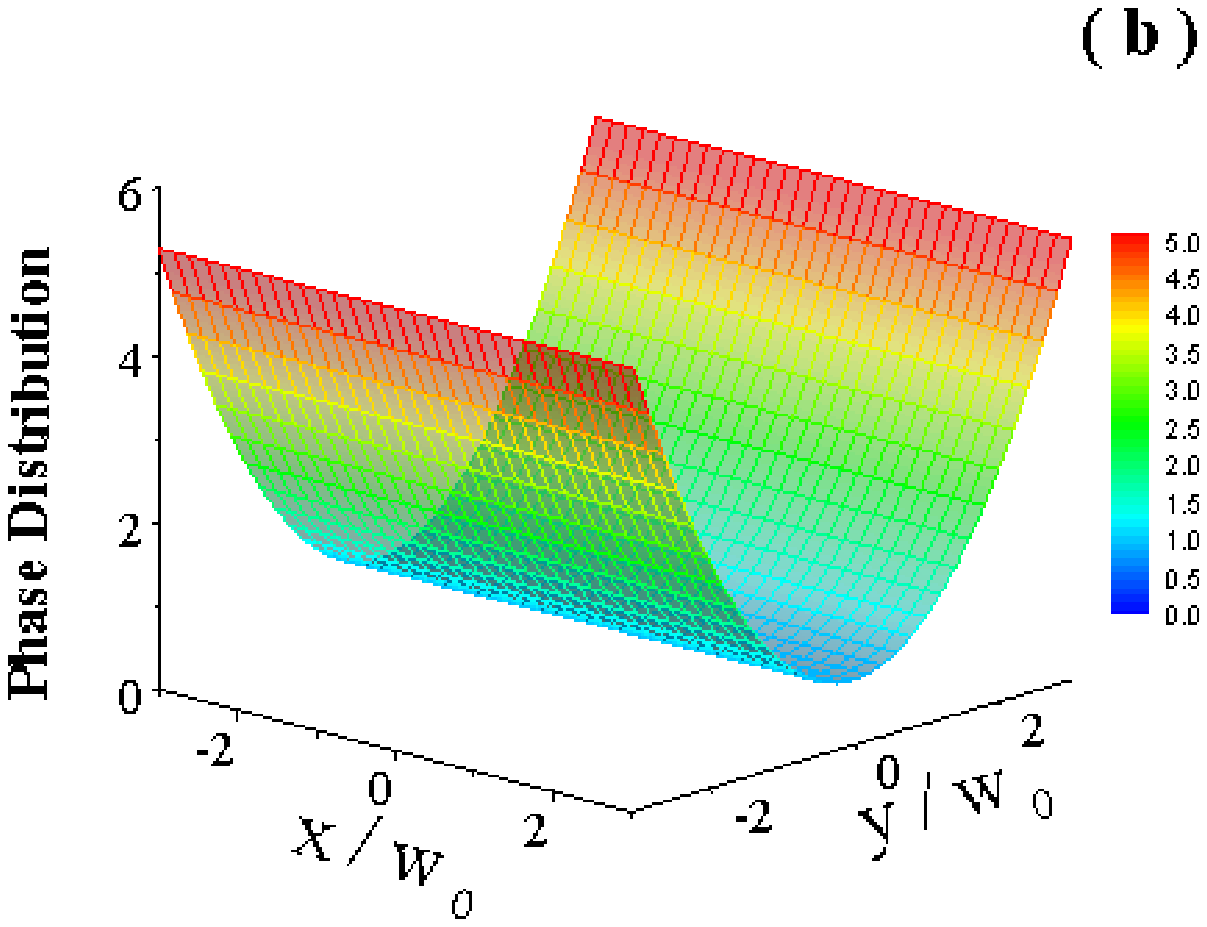}
\includegraphics[width=6cm]{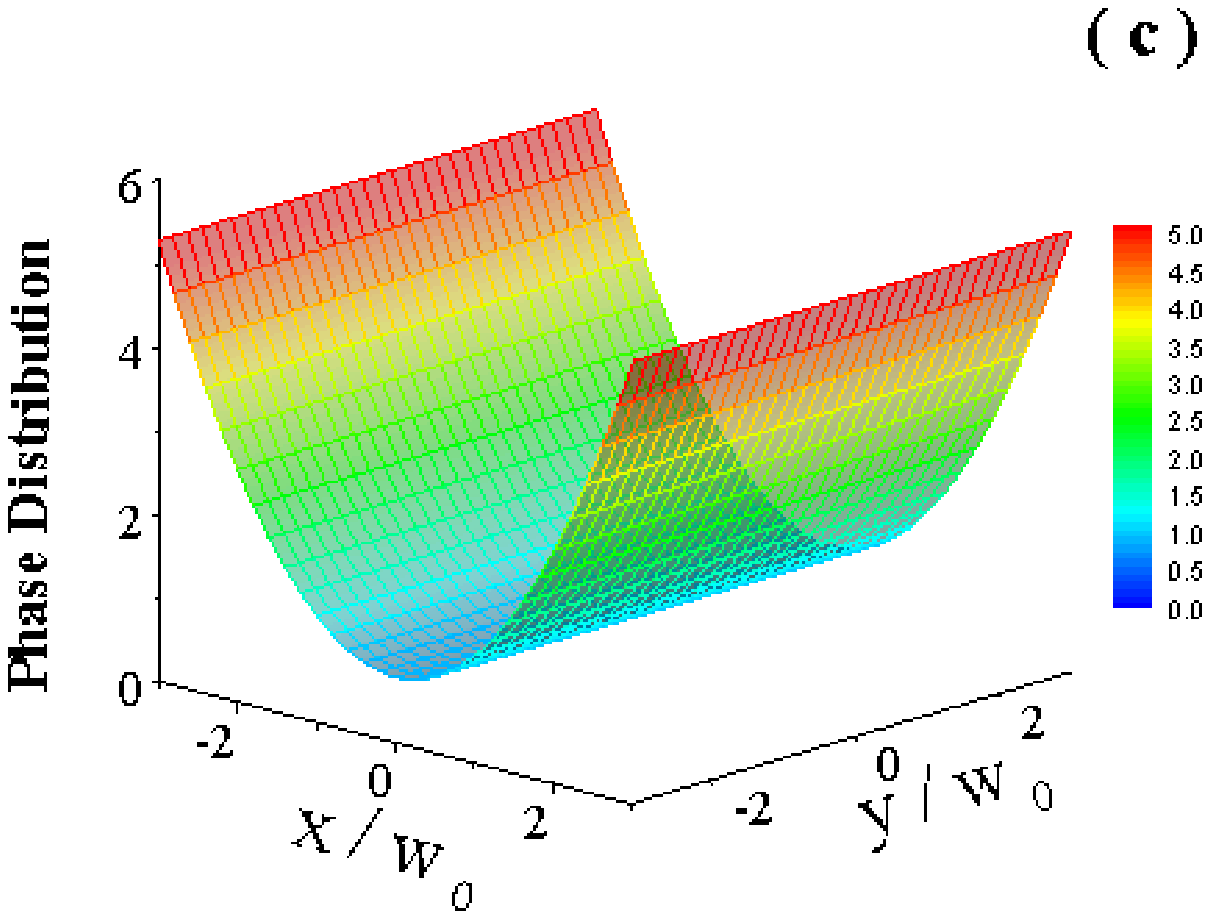}~~~~~
\includegraphics[width=6cm]{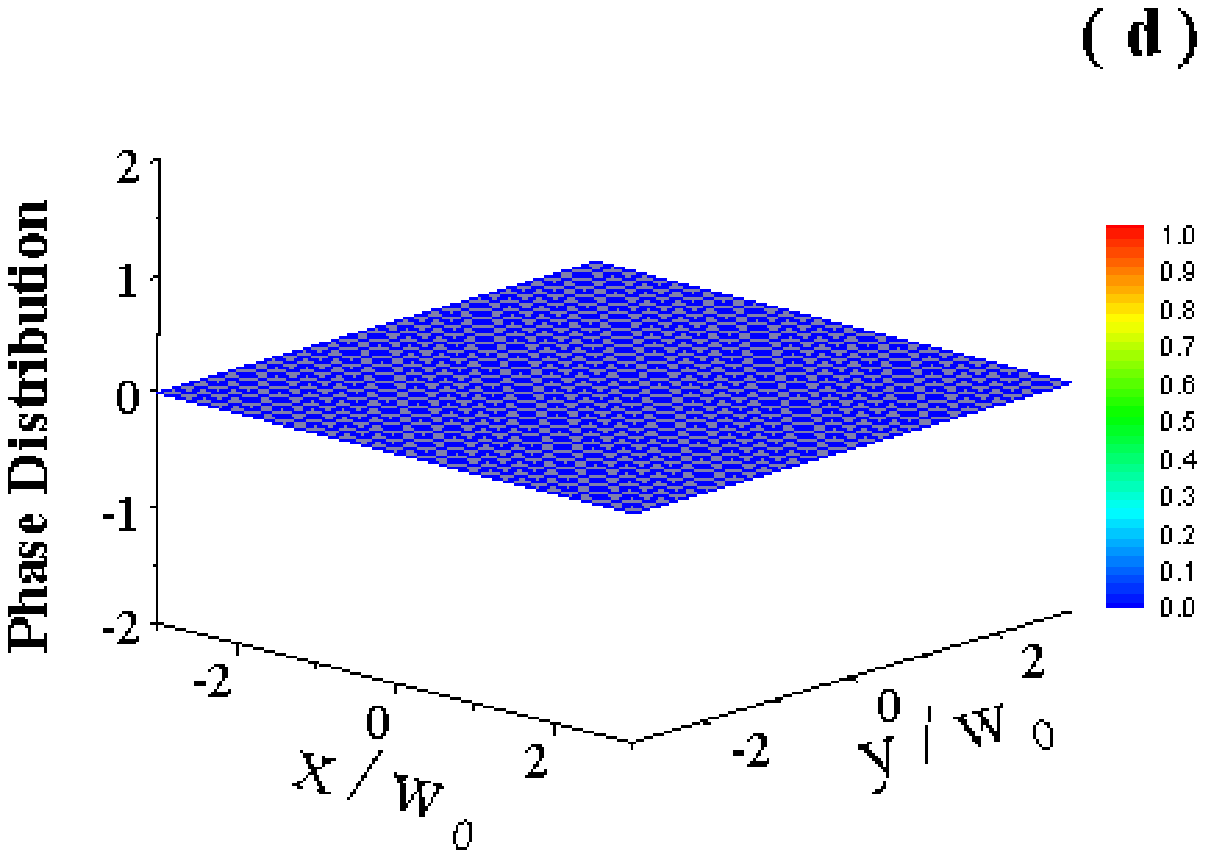}
% Here is how to import EPS art
\caption{\label{Fig7} (Color online) The numerically computed
phase distribution in object and image planes. (a) The phase
distribution in object plane. The phase distribution in image
plane after the Gaussian beam propagating through the QIMM slab
with different anisotropic parameters: (b) $n_x=-1$, $n_y=-2$. (c)
$n_x=-2$, $n_y=-1$.  (d) $n_x=-1$, $n_y=-1$. The phase
distribution can be completely reconstructed at the image plane.
The parameters are the same as in Fig.~\ref{Fig5}.}
\end{figure}

We might suspect whether the phase difference caused by the Gouy
phase shift in vacuum can be compensated by that caused by the
inverse Gouy phase shift in QIMM slab. To obtain the better
physical picture, the schematic distribution of phase fronts are
plotted in Fig.~\ref{Fig6}. The phase fronts of a focused Gaussian
beam are plotted with solid lines, and the phase fronts of a
perfect spherical wave are depicted with the dashed lines. The
phase difference on the optical axis is caused by the Gouy phase
shift. The inherent secret underlying the reverse Gouy phase shift
in the QIMM slab is the waves undergo a negative phase velocity.

Let us investigate what happens if we consider the phase
difference caused by the Gouy shift. Under the focusing matching
conditions, the phase difference caused by the Gouy phase shift in
the three regions are
\begin{eqnarray}
\Delta \Phi_1&=&-\arctan\frac{a}{z_R},\nonumber\\
\Delta \Phi_2&=& \arctan \frac{a}{z_R}+\arctan
\frac{b}{z_R},\nonumber\\
\Delta \Phi_3&=&-\arctan \frac{b}{z_R}.\label{gd}
\end{eqnarray}
The first and third equations dictate the phase difference caused
by the Gouy shift in regions $1$ and $3$, respectively. The second
equation denotes the phase difference caused by the inverse Gouy
phase shift in the QIMM slab. Subsequent calculations of
Eq.~(\ref{gd}) show
\begin{equation}
\Delta \Phi_1+\Delta \Phi_2+\Delta \Phi_3=0.\label{gouy}
\end{equation}
This implies that the phase difference caused by the Gouy phase
shift in vacuum can be compensated by the counterpart caused by
the inverse Gouy phase shift in QIMM slab. Therefore the condition
for phase compensation can be simply written as
\begin{equation}
(a+b)k_0+\sigma\sqrt{\varepsilon_y \mu_x} d k_0 =0.\label{pha}
\end{equation}
The first term in Eq.~(\ref{pha}) is the phase deference caused
by the plane wave in vacuum, and the other term is the phase
deference caused by the plane wave in the QIMM slab.

For the purpose of illustration, the phase distribution in object
plane is plotted in Fig.~\ref{Fig7}(a). Generally, the phase
distributions in image plane is distorted as shown in
Fig.~\ref{Fig7}(b) and Fig.~\ref{Fig7}(c). As mentioned above, the
phase distortion is caused by the effect of anisotropic
diffraction. To cancel the phase distortion, the beam waists
should locate at the same place, namely $z_{3x}^\ast=z_{3y}^\ast$.
Under the phase matching condition, the phase distribution at the
object plane can also be completely reconstructed at the image
plane as depicted in Fig.~\ref{Fig7}(d).

Now an interesting question naturally arises: whether the matching
conditions of focusing and the phase compensation can be satisfied
simultaneously. Clearly, if we seek a solution satisfying
Eqs.~(\ref{foc}) and (\ref{pha}), the only possibility is
\begin{equation}
\varepsilon_x=\varepsilon_y, ~~~\varepsilon_y \mu_z=1. \label{del}
\end{equation}
Under the matching conditions, the intensity and phase
distributions at the object plane can be completely reconstructed
at the image plane.

It should be mentioned that, for the QIMM slab with double-sheeted
hyperboloid wave-vector surface, both E- and H-polarized beams can
also exhibit the same intensity and phase reconstructed effect.
Because of the positive phase velocity embedded in this type of
QIMM, the paraxial beam will experience the positive Rayleigh
distance and Gouy phase shift. Therefore the accumulated phase
delay effect give rise to a large phase deference between the
object and image planes.

\section{The transmission of evanescent waves}\label{SecVI}
In this section, we discuss under what conditions anomalous
transmission will occur when an evanescent wave transmitted
through the QIMM slab. It is well known that when an evanescent
wave is transmitted through a slab of regular media with
simultaneously positive permittivity and permeability, the
amplitude of the transmitted wave will decay exponentially as the
thickness of the slab increases. While an evanescent wave
transmitted through an isotropic LHM slab, the amplitude of the
transmitted wave would be amplified exponentially. This anomalous
transmission of evanescent waves is a very peculiar property of
LHM and it may lead to subwavelength
image~\cite{Pendry2000,Fang2005}.

Now we will explore what happen when the evanescent wave
transmission through the QIMM slab. For E-polarized incident
waves, the incident and reflected fields in region 1 can be
written as
\begin{equation}
{\bf E} _{1} = E_0 {\bf e}_y \exp[i(k_x x+k_z z)]+R_E E_0 {\bf
e}_y \exp[i(k_x x-k_z z)],\label{E1}
\end{equation}
where $R_E$ is the reflection coefficient. Some of the incident
wave is transmitted into the QIMM slab, and conversely a wave
inside the QIMM slab incident on its interfaces with the
surrounding vacuum also experiences transmission and reflection,
so the electric field of the wave inside the slab is given by
\begin{equation}
{\bf E} _{2}  = r E_0{\bf e}_y \exp[i (q_x^{E} x+
q_z^{E}z)]+t E_0{\bf e}_y \exp[i (q_x^{E} x-  q_z^{E}z
)].\label{E2}
\end{equation}
Here $r$ and $t$ are coefficients which need to be determined by
boundary conditions~\cite{Hu2002}. Matching the boundary
conditions for each wave-vector component at the plane $z=a+d$
gives the propagation field in the form
\begin{equation}
{\bf E}_{3} = T_E E_0{\bf e}_y \exp[i q_x^{E} x+  q_z^{E}
(z-d)],\label{E3}
\end{equation}
where $T_E$ is the overall transmission coefficient. The
$z$-component of the wave vectors of evanescent waves can be found
by the solution of Eq.~(\ref{D2}), which yields
\begin{equation}
{q_z^E} = i \sqrt {\frac{\mu_x}{\mu_z}q_{x}^2-\varepsilon_y \mu_x
k_0^2}, ~~~{q_z^H} = i \sqrt
{\frac{\varepsilon_x}{\varepsilon_z}q_{x}^2-\varepsilon_x \mu_y
k_0^2}, \label{eqz}
\end{equation}
for E- and H-polarized waves, respectively.

\begin{figure}
\includegraphics[width=8cm]{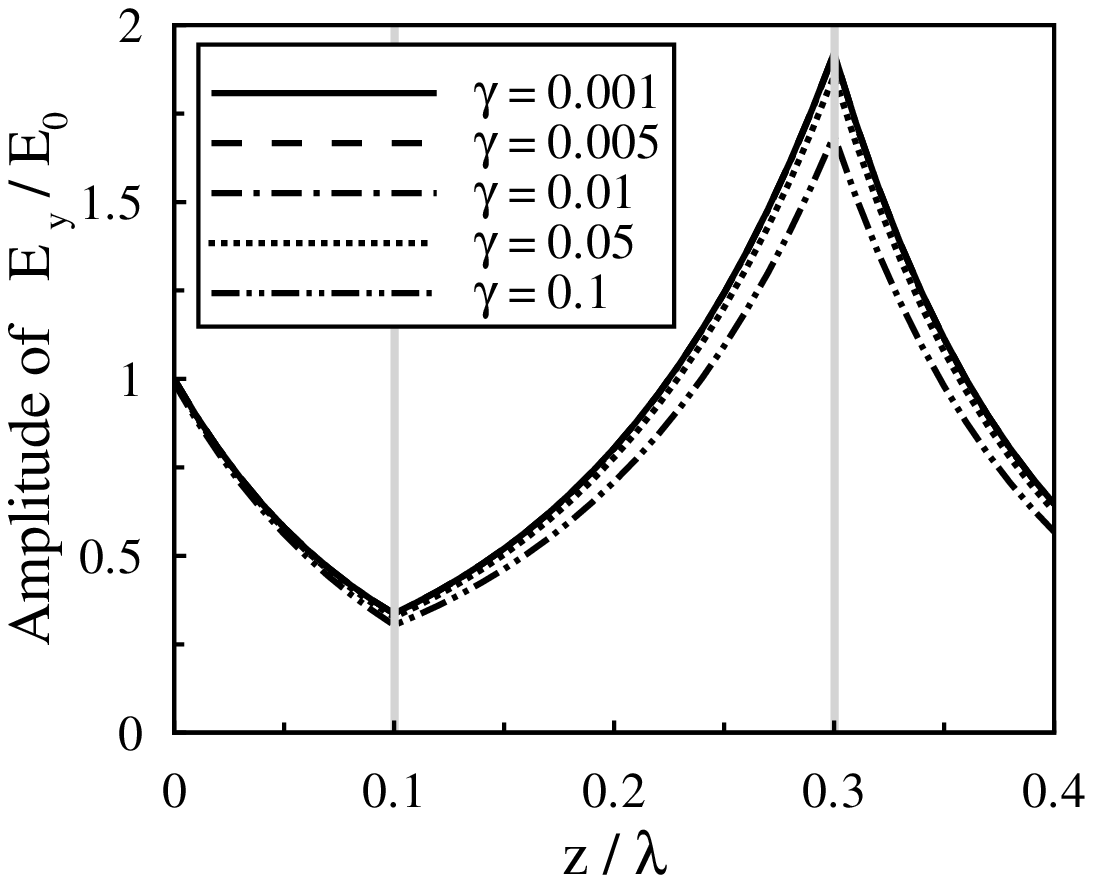}
\includegraphics[width=8cm]{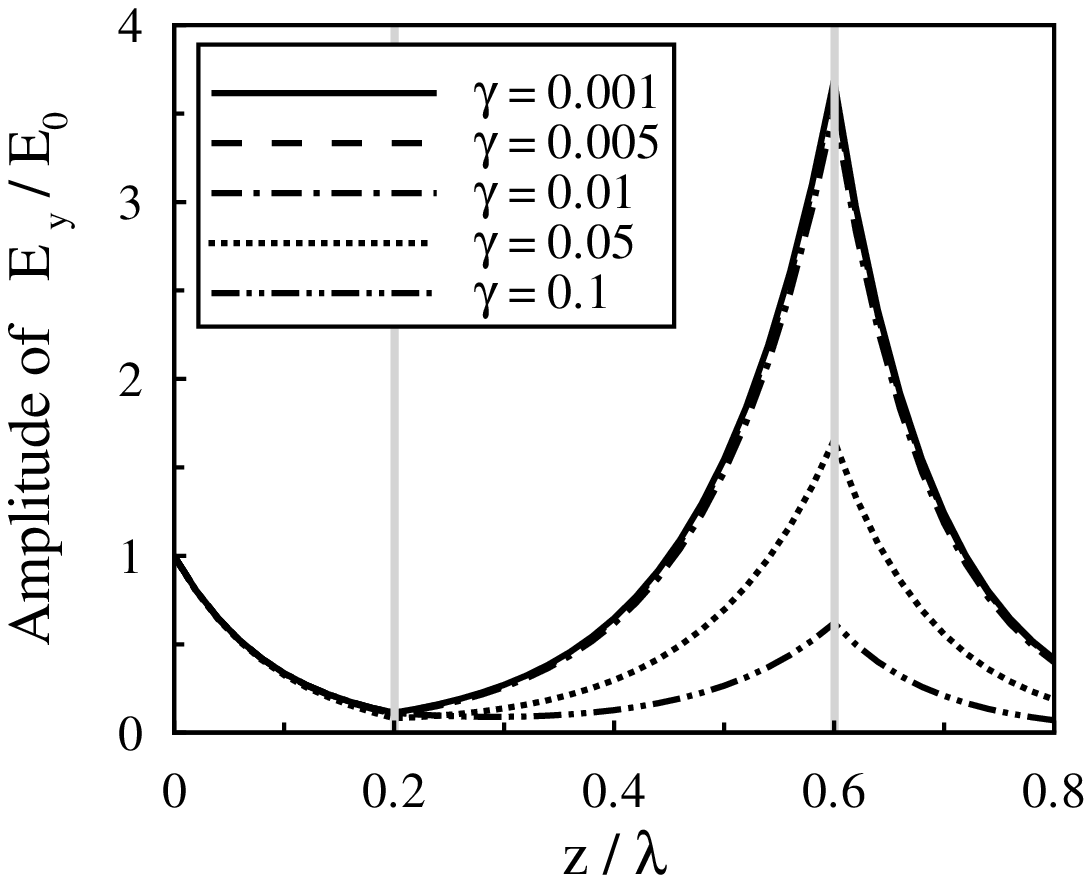}
% Here is how to import EPS art
\caption{\label{Fig8}  An evanescent wave ($k_x=2k_0$) interacting
with a QIMM slab  (a) $d= 0.2 \lambda$  (b) $d= 0.4 \lambda$. The
QIMM with different values of absorption
$\gamma_e=\gamma_m=\gamma$, ranging from 0.001 to 0.1. The QIMM
slab is supposed to be well satisfied the conditions:
$\varepsilon_x=\varepsilon_y=-0.8-\gamma i $ and
$\mu_z=-1.25-\gamma i$. The vertical gray lines denote the two
surfaces of the QIMM slab.}
\end{figure}

By matching the electric and magnetic fields at the two interfaces
between the QIMM slab and the surrounding vacuum, the coefficients
in Eqs.~(\ref{E1}), (\ref{E2}), and (\ref{E3}) can be determined.
We can get that the overall transmission through both surfaces of
the QIMM slab is given by
\begin{equation}
T_E =\frac{4 \mu_x k_z q_z^E \exp(iq_z^E d)}{(\mu_x
k_z+q_z^E)^2-(q_z^E-\mu_x k_z)^2 \exp(2iq_z^E d)}, \label{TE}
\end{equation}
From Eq.~(\ref{TE}) we can see that in general cases, when an
evanescent wave is transmitted through a QIMM slab, its amplitude
will decay exponentially as the thickness of the slab increases.
But if the following conditions are satisfied:
\begin{equation}
\mu_x<0,~~\mu_z< 0,~~\varepsilon_y<0,\label{EECI}
\end{equation}
\begin{equation}
\varepsilon_x \mu_z=1, ~~~\varepsilon_y \mu_z=1, \label{EECII}
\end{equation}
the amplitude of the transmitted evanescent wave will be amplified
exponentially by the transmission process through the QIMM slab.
From Eq.~(\ref{TE}), we can see that if the conditions
(\ref{EECI}) and (\ref{EECII}) are satisfied, the overall
transmission coefficient $T_E$ will be equal to $\exp(|q_z^E |d)$,
hence the amplitude of the transmitted evanescent wave will be
amplified exponentially as the thickness of the QIMM slab
increases. Note that the conditions mentioned above are sightly
different from those as Hu and Chui have obtained~\cite{Hu2002}.

As for H-polarized evanescent waves, the overall transmission
through both interfaces of the QIMM slab can be obtained by
similar procedures, and we can get  the overall transmission
coefficient:
\begin{equation}
T_H =\frac{4 \varepsilon_x k_z q_z^H \exp(iq_z^H
d)}{(\varepsilon_x k_z+q_z^H)^2-(q_z^H-\varepsilon_x k_z)^2
\exp(2iq_z^H d)}. \label{TH}
\end{equation}
From Eq.~(\ref{TH}), we can see that in general cases, the
amplitude of the transmitted H-polarized evanescent waves will
also decay exponentially as the thickness of the QIMM slab
increases. But if the following conditions are satisfied, the
overall transmission coefficient $T_H$ will be equal to
$\exp(|q_z^H| d)$, and the amplitude of the transmitted
H-polarized evanescent wave will be amplified exponentially by the
transmission process through the QIMM slab. So for H-polarized
evanescent waves,
\begin{equation}
\varepsilon_x<0,~~\varepsilon_z< 0,~~\varepsilon_y<0,\label{EHCI}
\end{equation}
\begin{equation}
 \varepsilon_z \mu_x=1, ~~~\varepsilon_z \mu_y=1.
\label{EHCII}
\end{equation}
if conditions (\ref{EHCI}) and (\ref{EHCII}) are satisfied, the
QIMM slab will enhance exponentially the transmitted waves.
Comparing (\ref{EECI}) and (\ref{EECII})  with (\ref{EHCI})  and
(\ref{EHCII}), we find that in the presence of QIMM, the
conditions for  E-polarized means automatically the conditions for
H-polarized are satisfied. Hence the QIMM also exhibit the
significant insensitive effect for evanescent waves. We stress
that, for the QIMM slab with double-sheeted hyperboloid dispersion
relation, the evanescent waves cannot be amplified.

In the above analysis the permittivity and permeability tensor
elements are assumed to be lossless. However the effect of
absorption, necessarily present in such materials, may drastically
suppress any evanescent amplifying wave into a decaying
one~\cite{Garcia2002,Rao2003}. Comparing Fig.~\ref{Fig8}(a) with
Fig.~\ref{Fig8}(b) suggests that large $d$ is not favored for the
amplification of the evanescent wave inside the QIMM slab. It is
also clearly seen from Figs.~\ref{Fig8}(a) and \ref{Fig8}(b) that
a low reflection occurs at the interfaces, since the QIMM slab is
supposed to be well satisfied the conditions (\ref{EECI}) and
(\ref{EECII}). So far we have shown the simulation results of the
evanescent wave interacting with QIMM slabs with different
absorptions and thicknesses. We find that the suppression of
evanescent-wave amplification can be effectively relaxed by
reducing the thickness of the QIMM slab. The numerical examples
provide direct evidence that an evanescent wave could be amplified
in a QIMM slab with finite absorption. Unfortunately, the loss of
a realistic metamaterial could not be reduced to a very small
level~\cite{Shen2003}. Therefore, the thickness of the QIMM slab
should be much smaller than the wavelength in order to realize the
subwavelength imaging for such a system.

As a result of the amplification of evanescent waves inside the
lossy QIMM slab, superlensing effect with subwavelength image
resolution could be achieved practically. While the main effect of
anisotropy associated with QIMM will limit the resolution of the
image. Although the image is imperfect, we trust that the
reconstruct effect of intensity and phase in paraxial regime will
lead to some potential applications. Several recent developments
make the polarization insensitive lens a practical possibility.
Some time ago it was shown that a double-periodic array of pairs
of parallel gold nanorods will exhibit negative permittivity and
permeability in the optical
range~\cite{Shalaev2005,Kildishev2006}. Another extremely
promising material has been previously explored in certain designs
of photonic crystals, which can be effectively modelled with
anisotropic permittivity and permeability
tensors~\cite{Shvets2003,Shvets2004,Urzhumov2005}. Experimentally,
the goal to realize the insensitive lens, chiefly lies in reducing
the loss of QIMM. Practical polarization insensitive lens will
require the frequency independent, a great challenge to the
designers is to realize the negative material parameters in a wide
band. The recent developments lead us to be optimistic that the
polarization insensitive lens can be designed in future.

\section{Conclusions}\label{SecVII}
In conclusion, we have proposed how to employ the QIMM slab to
create a polarization insensitive lens, in which both E- and
H-polarized waves exhibit the same refocusing effect. For shallow
incident angles the QIMM slab will provide some degree of
refocusing in the same manner as an isotropic negative index
material. We have investigated the focusing and phase compensation
of paraxial beams by the QIMM slab. We have introduced the
concepts of inverse Gouy phase shift and negative Rayleigh length
of paraxial beams in QIMM. We have shown that the phase difference
caused by the Gouy phase shift in vacuum can be compensated by
that caused by the inverse Gouy phase shift in the QIMM slab. If
certain matching conditions are satisfied, the intensity and phase
distributions at object plane can be completely reconstructed at
the image plane. The QIMM slab exhibits the significant
insensitive effect for both transmitted and evanescent waves.  Our
simulation results show that the superlensing effect with
subwavelength image resolution could be achieved in the form of a
QIMM slab. We wish the essential physics described in this paper
will provide reference in the road to construct the polarization
insensitive lens. We trust that the significant insensitive
properties will lead to further novel effects and applications.

\begin{acknowledgements}
H. Luo are sincerely grateful to Professor Thomas Dumelow for many
fruitful discussions. We also wish to thank the anonymous referees
for their valuable comments and suggestions. This work was
partially supported by projects of the National Natural Science
Foundation of China (Nos. 10125521 and 10535010), and the 973
National Major State Basic Research and Development of China (No.
G2000077400).
\end{acknowledgements}

\end{document}